\documentclass[twocolumn, twocolappendix]{aastex631}

\usepackage{amsmath}
\usepackage{soul}

\newcommand{\code}[1]{{\color{blue} \texttt{#1}}}

\newcommand{\oiii}{$\lbrack$OIII$\rbrack$}
\newcommand{\feii}{$\lbrack$FeII$\rbrack$}

\begin{document}

\title{Shocks, Winds, and a Torus: The Large Binocular Telescope Interferometer (LBTI) Resolves the Active Nucleus of NGC 4151}

\author[0000-0002-1272-6322]{Jacob W. Isbell}
\affiliation{Department of Astronomy and Steward Observatory, The University of Arizona, 933 N Cherry Ave, Tucson, AZ 85719, USA}

\author[0000-0002-2314-7289]{Steve Ertel}
\affiliation{Department of Astronomy and Steward Observatory, The University of Arizona, 933 N Cherry Ave, Tucson, AZ 85719, USA}
\affiliation{Large Binocular Telescope Observatory, University of Arizona, 933 N Cherry Ave, Tucson, AZ 85719, USA}

\author{Makoto Kishimoto}
\affiliation{Department of Astrophysics \& Atmospheric Sciences, Kyoto Sangyo University, Kamigamo-motoyama, Kita-ku, Kyoto 603-8555, Japan}

\author{Gerd Weigelt}
\affiliation{Max Planck Institute for Radio Astronomy, Auf dem Hügel 69, 53121 Bonn, Germany}

\author{J\"org-Uwe Pott}
\affiliation{Max Planck Institute for Astronomy, K\"onigstuhl 17, 69117 Heidelberg, Germany}

\author{Jared Carlson}
\affiliation{Department of Astronomy and Steward Observatory, The University of Arizona, 933 N Cherry Ave, Tucson, AZ 85719, USA}

\author{Qixiang Duan}
\affiliation{Department of Astronomy and Steward Observatory, The University of Arizona, 933 N Cherry Ave, Tucson, AZ 85719, USA}

\author{Violeta G\'amez Rosas}
\affiliation{Department of Astrophysics, Geophysics and Oceanography, University of Liege, Quartier Agora, allée du six Août 19c, 4000 Liège 1, Belgium}

\author{Walter Jaffe}
\affiliation{Leiden Observatory, Leiden University, Niels Bohrweg 2, NL-2333 CA Leiden, The Netherlands}

\author[0000-0001-6009-1803]{James Leftley}
\affiliation{Department of Physics \& Astronomy, University of Southampton, Southampton, SO17 1BJ, UK}

\author{Daniel May}
\affiliation{Instituto de Astronomia, Geofísica e Ciências Atmosféricas, Universidade de São Paulo, 05508-090, São Paulo, SP, Brazil}

\author{Romain. G. Petrov}
\affiliation{Laboratoire Lagrange, Universit\'e C\^ote d'Azur, Observatoire de la C\^ote d'Azur, CNRS, Boulevard de l'Observatoire, CS 34229, 06304 Nice Cedex 4, France}

\author{Jennifer Power}
\affiliation{Large Binocular Telescope Observatory, University of Arizona, 933 N Cherry Ave, Tucson, AZ 85719, USA}

\author{H\'el\`ene Rousseau}
\affiliation{Department of Astronomy and Steward Observatory, The University of Arizona, 933 N Cherry Ave, Tucson, AZ 85719, USA}

\author{Justin Rupert}
\affiliation{Large Binocular Telescope Observatory, University of Arizona, 933 N Cherry Ave, Tucson, AZ 85719, USA}



\begin{abstract}

We present mid-infrared (MIR) observations of the Seyfert 1 galaxy NGC 4151 using the Large Binocular Telescope Interferometer (LBTI).  
We took open-loop Fizeau images with 66--104 mas (5.8--9.1 pc) resolution in the N-band (at $8.7$ and $10.5$~\micron), using the full resolution of the LBTI -- equivalent to that of a 28.8 m telescope. These images were complemented by AO imaging in the LM-bands ($3.7$ and $4.8$~\micron), with 50--62 mas (4.4--5.4 pc) resolution.
These images bridge the scales between previous Very Large Telescope Interferometer (VLTI)/MIDI and VLT/VISIR data, delivering ELT-like imaging resolution in the N-band. 
We resolve a dusty torus, (diameter 32 pc, PA$=125^{\circ}$), and detect dusty clouds within the narrow line region. 
Matching the resolution across four bands, we measured spatially-resolved SEDs of the central $\sim 100$~pc. 
Modified blackbody fitting revealed dust temperature and extinction profiles, indicating both heating from the accretion disk and additional shock heating due to the radio jet. 
The spatial coincidence of ionized emission (e.g., \feii~and \oiii), extended MIR structures, and radio features further supports the interpretation of shock heating. 
Comparison with NGC 1068 tests the Unified Model of Active Galactic Nuclei (Unified Model of AGN): Structures are similar, despite differences in orientation and Eddington ratio. NGC 4151’s torus is smaller than NGC 1068’s following a $r\propto L^{0.5}$ scaling. 
These ELT-like observations of NGC 4151 and NGC 1068 highlight the need to revise MIR radiative transfer models of AGN to account for jet-related heating. 
\end{abstract}



\section{Introduction}
\label{sec:intro}
The Unified Model of Active Galactic Nuclei (Unified Model of AGN) asserts that the observed differences between Seyfert 1 and Seyfert 2 (Sy1 and Sy2) AGN are due to the orientation of some obscuring structure relative to us. For many years this structure was thought to be a dusty molecular torus \citep[e.g.,][]{antonucci1993, nenkova2008}, but high-resolution studies with the Very Large Telescope Interferometer (VLTI) instruments MIDI \citep{leinert2003} and MATISSE \citep{lopez2022} have given strong evidence of a structure consisting of a thin disk and polar-oriented dusty winds in both Sy1s \citep{hoenig2012, hoenig2013} and Sy2s \citep[][]{burtscher2013, tristram2014, isbell2022, isbell2023, gamezrosas2022}.  Models of a geometrically thin, optically thick disk and a bi-cone representing a dusty wind (disk+wind models) are also strongly supported by SED fitting to various AGN \citep[][]{honig2017, stalevski2019, isbell2021, garcia-bernete2022} and hydrodynamical modeling \citep{wada2016, williamson2020}. So far, high-resolution mid-infrared (MIR) model-independent images of the circumnuclear structures exist only for two Sy2 AGN and no Sy1 AGN. 

The prototypical Sy2, NGC 1068, was imaged at sub-parsec scale with VLTI/MATISSE \citep{gamezrosas2022} and at 5 pc resolution with the Large Binocular Telescope Interferometer (LBTI). The LBTI images \citep{isbell2025}, though at lower resolution, provided simultaneous insights into the optical depth of the thin disk, the extent of the dusty wind, and secondary heating processes related to the radio jet rather than the accretion disk. The images, at similar resolution to future extremely large telescopes (ELTs) at MIR wavelengths, supported the disk+wind flavor of the Unified Model of AGN, but also emphasized the importance of considering shock heating from the radio jet, even though NGC 1068 is a relatively radio-quiet AGN. Whether the Unified Model of AGN applies to other Seyferts remains to be tested, so this work uses similar LBTI imaging methods and resolution to do a comparative study in the Sy1 AGN, NGC 4151.

NGC 4151 is the closest Sy1 galaxy at a distance of between $15.8\pm0.41$~Mpc \citep{yuan2020} and $19.9_{-2.6}^{+2.5}$~Mpc \citep{honig2014}\footnote{Throughout, we adopt a distance to NGC 4151 of 18 Mpc, which is the average of the two presented measurements.}.  
According to the Unified Model of AGN, the primary difference between NGC 4151 and a Sy2 AGN (e.g., NGC 1068) should be orientation of the nuclear structures. This difference has so far been supported by the fact that very hot ($1500$~K) dust in the sublimation zone has been directly observed in NGC 4151 \citep{kishimoto2022} but not in NGC 1068 \citep[][]{gravity2020, gamezrosas2022} where it is instead heavily obscured by the optically thick disk. 

The LBTI \citep{hinz2016, ertel2020, isbell2024} co-phases and interferometrically combines the beams from the two 8.4~m mirrors of the Large Binocular Telescope (LBT), separated 14.4~m.  In Fizeau imaging, this results in a virtual telescope aperture with a resolution equivalent to that of a 28.8 m telescope (interferometric resolution is $\lambda / 2B$, where $B$ is the center-to-center separation of the apertures). Unlike other long-baseline interferometers, LBTI Fizeau imaging is \textit{direct imaging}, and does not rely on image reconstruction from visibilities and phases. This also means that it does not resolve-out structures like the VLTI does, resulting in high dynamic range.
Open-loop Fizeau imaging -- where the interferometric phase between the two apertures is not actively controlled -- is the simplest interferometric mode of the LBTI, and it can bridge the gap between high-dynamic-range single-dish observations with, e.g., JWST, and low-dynamic-range high-resolution interferometric images with, e.g., VLTI and the Center for High Angular Resolution Astronomy (CHARA) array. LBTI's 28.8 m effective aperture is directly comparable to the upcoming generation of $30$~m class telescopes, and it can serve as a key testing ground for science cases and techniques in the MIR. 

In this paper, we study the nearby Sy1, NGC 4151, using N-band open-loop Fizeau imaging and LM-band adaptive optics (AO) imaging at the LBTI. These images provide spatial constraints on the structure and heating of the nuclear outflow and dust torus. We compare these results to previous imaging of NGC 1068 to test the Unified Model of AGN.

\begin{table*}[t]
    \centering
    \hspace{-3cm}
    \begin{tabular}{l|cccc|rlr}
         Source  & Filter & $\lambda_c$ $\lbrack\mu$m$\rbrack$ & Obs. Type & Date & N$_{\rm exp}$ & Exp. Time [s] & Field Rotation [deg] \\\hline
         NGC 4151& Std-L & 3.7 & Single-dish & 2025-01-11 & 4800 &0.3571 & 7$^\circ$\\
         NGC 4151& Std-M & 4.8 & Single-dish & 2025-01-11 & 6400 & 0.206& 9$^\circ$\\ 
         NGC 4151& W\_08 & 8.7 & Fizeau & 2024-04-21 & 116000 & 0.0427 & 136$^\circ$ \\ 
         NGC 4151& W\_10 & 10.5 & Fizeau & 2024-04-22 & 96000 & 0.0213 & 140$^\circ$\\
         \hline
         NGP 39 74& Std-L & 3.7 & Single-dish & 2025-01-11 &3200 &0.3571 &-\\
         NGP 39 74& Std-M & 4.8 & Single-dish & 2025-01-11& 3200 &0.206 &-\\ \hline
         HD 105140& W\_08 & 8.7 & Fizeau & 2024-04-21 & 44000 & 0.0427 &-\\ 
         HD 105140& W\_10 & 10.5 & Fizeau & 2024-04-22 & 44000 & 0.0213 &-\\
    \end{tabular}
    \caption{LBTI observations entering this work. Exposure time is given per frame, and the number of exposures is N$_{\rm exp}$.}
    \label{tab:observations}
\end{table*}

\section{Observations and Data Reduction}
We obtained infrared observations of the Sy1 galaxy, NGC 4151, using the Large Binocular Telescope Interferometer (LBTI). The observations were carried out using both the LBTI/LMIRCam camera \citep{leisenring2012} and the LBTI/NOMIC camera \citep{hoffmann2014}. The observations with LBTI/NOMIC were done in open-loop Fizeau imaging mode, utilizing the resolving power of the LBTI's 28.8m effective aperture. The observations with LBTI/LMIRCam were standard AO-assisted images using one of the 8.4~m apertures of the LBT. Details of the observations are given below. We list the targets, observing modes, filters, and exposure times of our observations in Table \ref{tab:observations}.

\label{sec:obs}
\subsection{Fizeau Observations}
Using LBTI/NOMIC we observed NGC 4151 and a point-spread-function (PSF) calibrator, HD 105140, on the nights of 21 April and 22 April 2024 in open-loop Fizeau imaging mode. In the first night we used the $8.7~\mu$m filter, obtaining $136^{\circ}$ of field rotation and 4782.4 seconds of total exposure. In the second night we used the $10.5~\mu$m filter, obtaining $141^{\circ}$ of field rotation and 2002.2 seconds of total exposure. We attempted to start the observations at similar initial parallactic angles so that the resulting field rotation would be as similar as possible in both filters. In both nights, we utilized an A-B nodding pattern, exposing for 2000 frames in position A and then nodding $2"$ to position B for another 2000 frames. This was repeated throughout the night, interrupted only by intermittent observations of the PSF and flux calibrator, HD 105140. These observations were taken in the LBTI open-loop mode, meaning that we utilize lucky fringing to ensure our final observations are properly co-phased. In order to maximize the number of ``lucky'' frames, we performed manual, slow co-phasing of the apertures using simultaneous, spectrally dispersed LMIRCam images. 
The frame selection criteria are described in \S\ref{sec:reduction}.

\begin{figure*}
    \centering
    \includegraphics[width=0.9\linewidth]{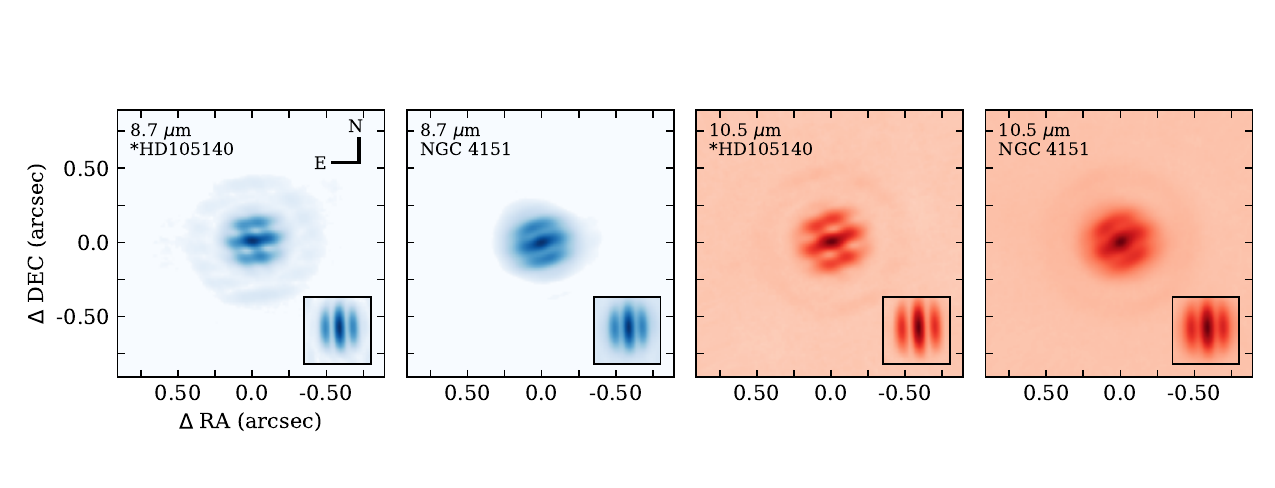}
    \vspace{-0.8cm}
    \caption{LBTI/NOMIC Fizeau images of NGC 4151 and PSF calibrator HD105140. The left two panels (in blue) show the 8.7 \micron~images of the sources after frame selection, corotation, and stacking. The right two panels (in red) show the same at 10.5 \micron. The PSF calibrator has been artificially rotated to match the observed position angles of the target. Inset in each panel is the median Fizeau PSF without corotation; it shows the characteristic fringe pattern and the quality of the frame selection. Extended emission is immediately visible in the NGC 4151 images; the images are much more filled in with extended flux than the calibrators.}
    \label{fig:psfs_fizeau}
\end{figure*}

\begin{figure*}
    \centering
    \includegraphics[width=0.9\linewidth]{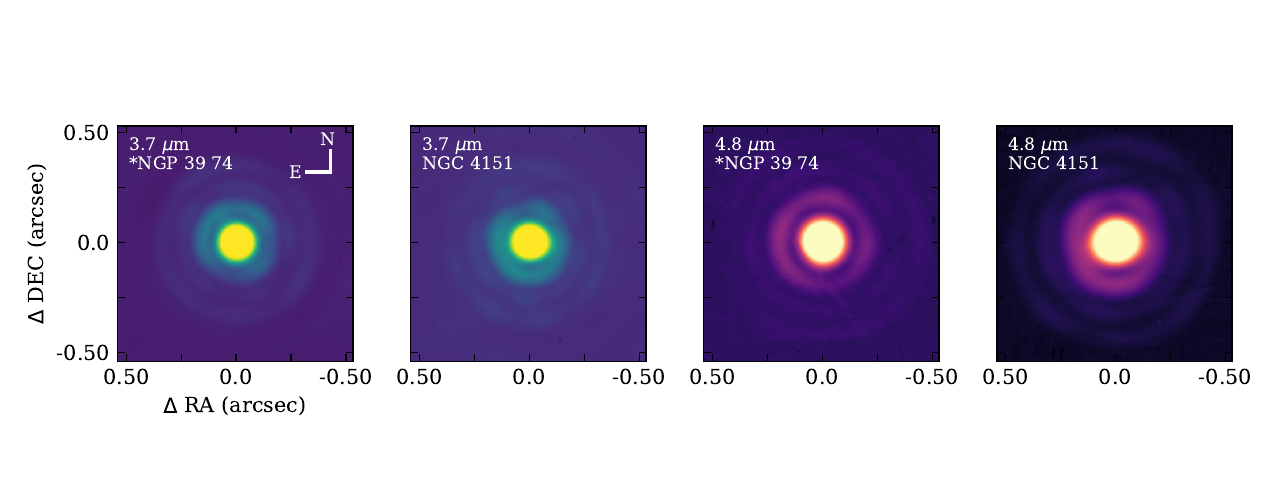}
    \vspace{-1cm}
    \caption{LBTI/LMIRCam AO images of NGC 4151 and PSF calibrator NGP 39 74. The left two panels (in viridis) show the 3.7 \micron~images of the sources after frame selection, corotation, and stacking. The right two panels (in magma) show the same at 4.8 \micron. The PSF calibrator has been artificially rotated to match the observed position angles of the target. Extended emission is visible near the cores of the NGC 4151 images relative to the calibrator.}
    \label{fig:psfs_ao}
\end{figure*}

\subsection{Single-Dish Observations}
Using LBTI/LMIRCam we observed NGC 4151 and a PSF calibrator, NGP 39 74, on the night of 11 January 2025 in standard AO imaging mode. We used the Std-L ($\lambda_c = 3.7~\mu$m) and Std-M ($\lambda_c = 4.78~\mu$m) filters. We obtained 1714.1 and 1318.4 seconds of total exposure, respectively. 
Similar to the Fizeau images, we utilized an A-B nodding pattern, exposing for 800 frames in position A and then nodding $8"$ to position B for another 800 frames.

\subsection{Data Reduction}
\label{sec:reduction}
Data processing for the Fizeau and single-dish images followed the procedure developed by \citet{isbell2024, isbell2025}. All processing steps are completed using the \code{LIZARD} pipeline developed for the LBTI\footnote{\href{https://github.com/jwisbell/lbti_fizeau}{https://github.com/jwisbell/lbti\_fizeau}}. 

First the A-B nods were subtracted to remove the bright MIR background. Each frame can then be flux calibrated using the $LMN$-band fluxes of the calibrators from \citet{cruzalebes2019} to estimate the LBTI filter fluxes: we used HD105140 for the $N$-band and NGP 39 74 for the $LM$-bands. Flux uncertainties are given from variations in total image flux in each frame. The calibrator flux uncertainty is added in quadrature. The total relative flux error is used to get per-pixel uncertainties, assuming noise is consistent across the detector. Relative uncertainties are typically slightly larger than $10\%$, primarily driven by the calibrator flux uncertainty. 

Second, ``good'' Fizeau frames were selected from the lucky fringing exposures. This step is not necessary for the single-dish images, and it is skipped for those images in the pipeline.  
Coherently interfered fringes are those with a relative phase shift smaller than one wavelength. Frames with good fringes are selected using the method described by \citet{isbell2024}. While in principle a large fraction of the frames could be kept ($\geq 90\%$ for this bright source) we opted to keep only the best $10\%$ of frames in order to consistently compare our results to those for NGC 1068 \citep{isbell2025}. This sacrifices some image sensitivity, but results in increased PSF stability.

Third, each frame was rotated so that North is up and East is left. Then the co-rotated images were stacked to produce the final science images. While each individual Fizeau imaging exposure only exploits the 29~m PSF along one baseline, using field rotation one can achieve high resolution at all image orientations \citep{isbell2024}. Following flux calibration and co-rotation of the science target images, the same set of rotations is applied to the median of the PSF calibrator exposures. This gives an empirical PSF estimate in each filter in both Fizeau and standard observing mode. The final Fizeau PSF estimate has a FWHM of $66.6\times104.4$ mas ($5.8\times9.1$~pc) and $84.6\times104.4$ mas ($7.5\times9.1$~pc) at 8.7 and 10.5~$\mu$m, respectively. The single-dish PSF FWHM is  $50.3\times50.3$~mas ($4.4\times4.4$~pc) and $62.1\times62.1$ mas ($5.4\times5.4$~pc) at 3.7 and 4.8 $\mu$m, respectively. 
The stacked+co-rotated science images and PSF estimates are shown in Figs. \ref{fig:psfs_fizeau} and \ref{fig:psfs_ao}.

\subsection{Image Deconvolution}
Using the stacked+co-rotated science images and the empirical PSF estimates, we perform PSF deconvolution to recover the underlying flux distribution of the science target in each filter. \citet{isbell2025} showed that both Richardson-Lucy \citep[R-L,][]{richardson1972, lucy1974} deconvolution and a CLEAN-inspired method recover the flux distribution with high fidelity. We employ both methods here because they each have specific strengths. \citet{isbell2024} showed that the CLEAN-inspired method better recovers low-surface brightness features. The R-L method, however, does not rely on the user-defined choice of a restoring beam. We show the CLEAN deconvolution results in Fig. \ref{fig:deconv} and the R-L results in Appendix Fig. \ref{fig:deconv-rl}. The R-L deconvolutions produce generally more compact structures, but the general shape and scale of the R-L results nonetheless match the CLEAN results.

\begin{figure*}
    \centering
    \includegraphics[width=0.99\linewidth]{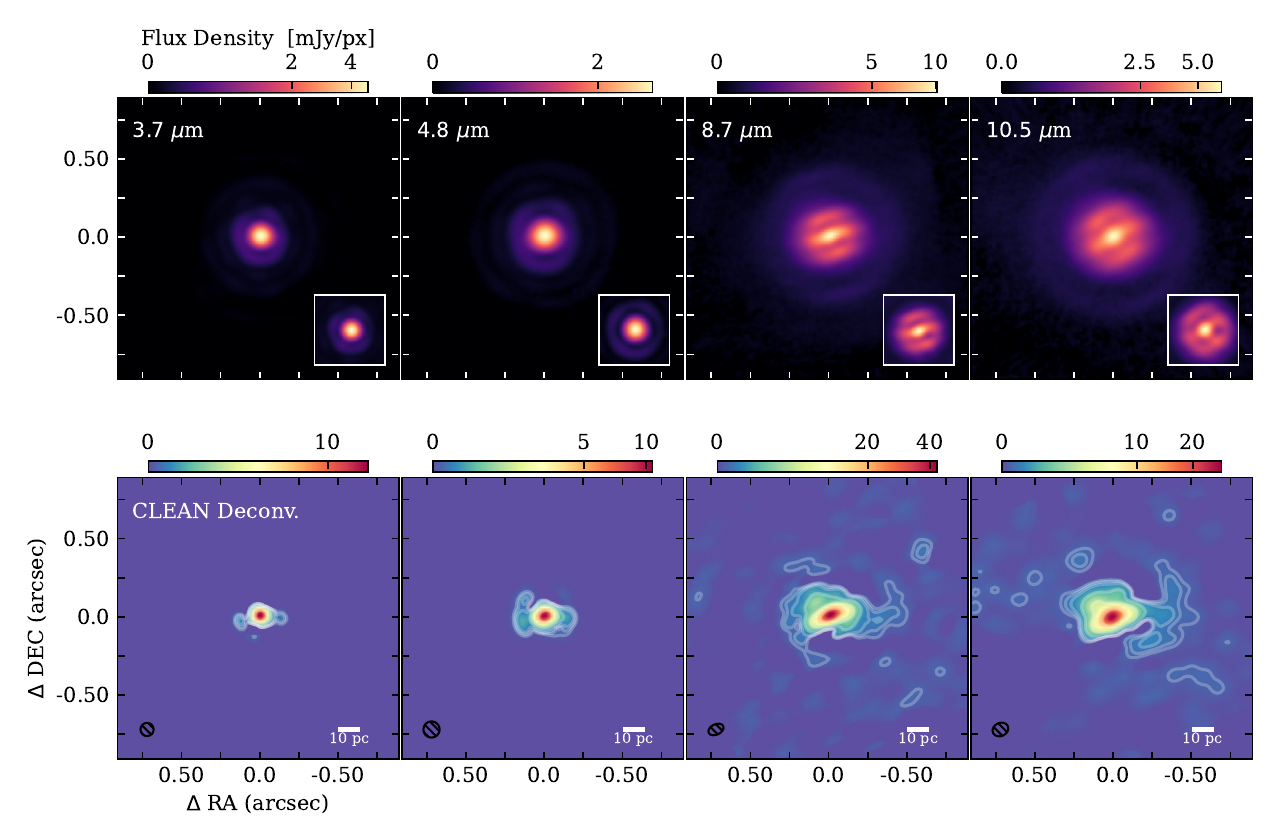}
    \caption{Deconvolved images of the nucleus of NGC 4151. \textit{Top row}) Stacked, corotated images at each wavelength (as in Figs. \ref{fig:psfs_fizeau} and \ref{fig:psfs_ao}) with PSF calibrator inset. 
    \textit{Bottom row}) CLEAN deconvolution results at each wavelength. The restoring beam of each image is given in the lower left. 
    Contours start at $95\%$ of the peak flux and decrease by factors of 2 down to a factor of 512. 
    The LMIRCam images have been rescaled to match the pixel scale of the NOMIC images (18 mas/px).}
    \label{fig:deconv}
\end{figure*}

CLEAN deconvolution requires the choice of the number of iterations, the gain, and the restoring beam \citep{hogbom1974}. In our case, we estimate the restoring beam (size and PA) from the FWHM of the empirical PSF. The gain was set to 0.05 for relatively fast convergence. 
Due to the extended nature of this source, we added the \code{phat} parameter as included in the National Radio Astronomy Observatory (NRAO) Common Astronomy Software Applications (CASA) version of multi-scale CLEAN. This parameter sets the relative strength of a spike at the center of the subtracted PSF, and it is used to better recover extended flux \citep{cornwell1983}. The restoring beams and CLEAN deconvolution parameters for each filter are given in Table \ref{tab:clean}.  

\begin{table}[]
    \centering
    \begin{tabular}{l|ccc|rr}
          $\lambda_c$ & n\_iter & gain & phat & FWHM & PA  \\
          $\lbrack$\micron$\rbrack$ & & & & $\lbrack$mas$\times$mas$\rbrack$ & $\lbrack$$^{\circ}$$\rbrack$ \\
         \hline
          3.7 & $5\times10^4$ & 0.05 &0.5&$50.3\times50.3$& 0 \\
          4.8 & $1\times10^5$ &0.05 &0.5&$62.1\times62.1$& 0 \\
         \hline
          8.7 & $8\times10^5$&0.05 &0.5&$66.6\times104.4$& -65\\
          10.5 & $4\times10^5$ & 0.05&0.5&$84.6\times104.4$&-65 \\
    \end{tabular}
    \caption{CLEAN deconvolution parameters. Position angle (PA) is the direction of the major axis.}
    \label{tab:clean}
\end{table}

In addition, we used the \code{scikit-image} Python package implementation of R-L deconvolution. This required the selection of a number of iterations (\code{niter}) and a cutoff for faint features to avoid issues in division (\code{eps}). We have balanced these values to maximize the extended features without causing obvious artifacts (such as a square feature around the edge of the image). The values are \code{niter=128} and \code{eps=1e-2}. 
 
Both approaches give very similar results, but the CLEAN implementation recovers more low surface brightness features, as previously shown for NGC 1068 \citep{isbell2025}. Because the total image flux is conserved in both approaches, the more compact R-L structures exhibit higher flux densities. For the remainder of this work, we focus on the CLEAN-deconvolved images.

\section{SED Fitting}
By performing spatially resolved SED fitting on the LBTI images of NGC 4151, we determined the temperature profile and emission characteristics of the central $\sim100$ pc. The deconvolved images were first matched in resolution by applying the same CLEAN restoring beam to each image. The restoring beam was selected to match that of the lowest resolution image: the 10.5 \micron~deconvolved Fizeau image. The images are then cross-matched such that the peak flux of each image is cospatial with the others'. This allowed us to extract per-pixel ($18\times 18$ mas) SEDs across four wavelengths. Per-pixel uncertainties on the extracted fluxes are propagated from the total flux relative uncertainties described above. Fluxes that are less than 2$\sigma$ above the  standard deviation of the CLEAN residuals (averaged across the image) are considered upper limits.

We fit a modified blackbody to each pixel's SED with the form
\begin{equation}
    F_{\lambda}(T, A_{\rm V}) = B_{\lambda}(T)~e^{\frac{-A_{\rm V, eff}}{1.09} \times \frac{\kappa_\lambda}{\kappa_{0.5}} },\label{eq:bb}
\end{equation}
where $T$ is the temperature in Kelvin, $A_{\rm V, eff}$ is the extinction/emission due to Silicates along the line of sight normalized to 0.5 \micron (see App. \ref{app:sed_derivation} for details on the approximation), and $\kappa_\lambda/\kappa_{0.5}$ is the mass extinction coefficient (normalized to 0.5 \micron) from the standard interstellar medium (ISM) $\kappa_\lambda$ profile given in \citet{schartmann2005} which is based on the standard ISM size distribution and composition of \citet{mathis1977}. The Planck function $B_{\lambda}(T)$ is integrated over each square pixel ($18\times18$~mas) to give a flux density in Jy.

$A_{\rm V,eff}$ is allowed to have negative values (i.e., negative extinction), which indicate that the Silicate feature is found in emission rather than absorption. In Sy2s the silicate feature has been found in absorption with MATISSE \citep[e.g.,][]{gamezrosas2022, isbell2022, isbell2023}, but it is found in emission in Young Stellar Objects (YSOs) \citep[e.g.,][]{varga2025}. In our simple model, the emission/absorption is assumed to come from a single foreground screen, and it neglects non-LTE effects such as scattering or multiple foreground absorbers. Full radiative transfer in future work would be necessary to provide more robust estimates of dust temperature and composition.

We fit the modified blackbody to the per-pixel SEDs using a brute-force grid search. Taking upper limits into account, we compute the $\chi^2$ value for each combination of temperatures, silicate feature strength, and emissivities. Temperatures were sampled every 5 K in the range $T \in [150, 900]$~K. The silicate feature strength was sampled every 1 mag in the range $A_{\rm V}\in [-100,120]$~mag. By searching the resulting $\chi^2$ grid, parameter uncertainties were defined using the typical $\chi_{min}^2+1$ estimator. Some example SED fits are shown in Appendix Fig. \ref{fig:sed-fit-examples}.

The fitted blackbody color temperatures and extinctions are given in Fig. \ref{fig:color_temp}. Uncertainties on the fitted values are given in the Appendix. We find that the region immediately next to the AGN shows the hottest temperatures, as expected. The temperatures generally decrease with distance from the nucleus, but there are regions of higher temperature both to the NE of the nucleus and far to the west. These are discussed below.

\begin{figure*}
    \centering
    \includegraphics[width=0.9\linewidth]{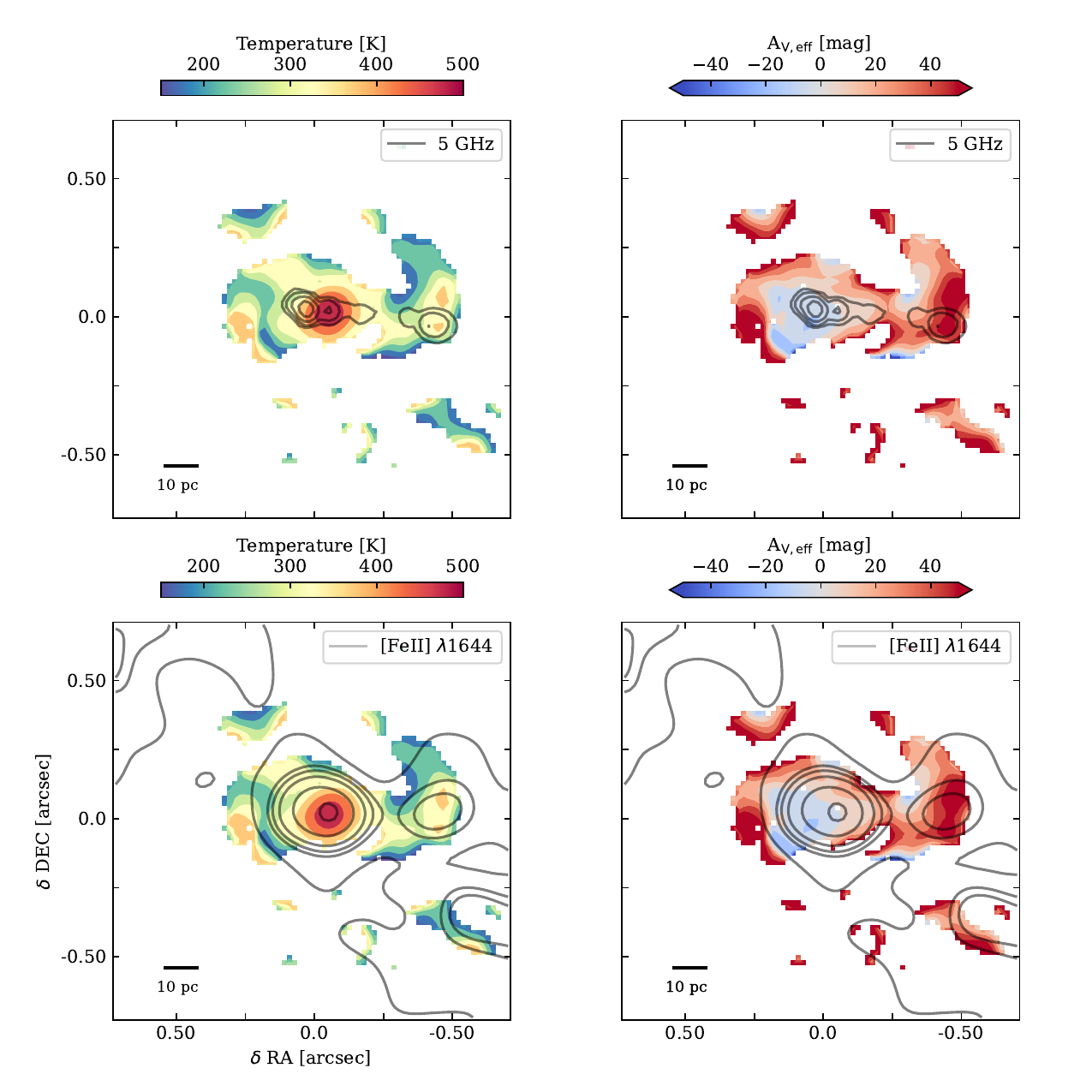}
    \caption{Fitted modified blackbody temperatures and extinction values for the MIR emission in NGC 4151. \textit{Top row}) In the \textit{left} panel, the 5 GHz radio contours \citep{williams2020} are overplotted on the temperature profile. In the \textit{right} panel, the same contours are plotted over the effective A$_{\rm V}$ profile. Negative values indicate silicate emission rather than absorption. \textit{Bottom row}) same as the top row but contours from the \feii~emission \citep{may2020}. There is a visible increase in temperature coincident with the \feii~cloud, immediately north of a radio knot.}
    \label{fig:color_temp}
\end{figure*}

\section{Discussion}
\label{sec:discussion}
The LBTI MIR images at 50--104 mas (4.4-9.1 pc) resolution reveal complex circumnuclear dust structures in NGC 4151. Four independent filters allowed us to construct per-pixel temperature maps. Below, we discuss the morphology of the circumnuclear dust and its relation to the ``dusty torus'' of the Unified Model of AGN. By comparing our images of NGC 4151 (Sy1) to previous results for NGC 1068 (Sy2), we test the Unified Model of AGN. We also discuss the temperature map and its indications of shock heating by the radio jet. This analysis is aided by comparisons to previously published observations in the radio \citep[5 GHz;][]{williams2020} and Visible/NIR \citep[\oiii$\lambda501$, H$\alpha \lambda656$, \feii$\lambda1644$, and H$_2 \lambda 1748$;][]{hutchings1999,may2020}.

\subsection{Alignment of Different Observations}
Naively, we could place the SMBH at the peak of the MIR emission in our images.   
However, \citet{isbell2025} found, through cross-correlation with the \oiii~emission, that the SMBH was located 50 mas south of the MIR peak in NGC 1068. In this work, we use previous kinematic arguments from \citet{may2020} to locate the SMBH. Those authors found that the \feii~emission peak matched the $\lbrack$Si VII$\rbrack$ kinematic center, and thus they identified it with radio feature C4. In later work, the SMBH was shown to reside in the radio feature C4W  \citep{williams2020}; we therefore place C4W at the \feii~peak. Finally, we cross-correlate the \feii~emission and our MIR images to align them astrometrically. Numerous features are found in both the MIR and \feii~images (i.e., the western arc, a central bar, and various low surface-brightness clouds). The same position is found if we instead cross-correlate with the \oiii~image. This places C4W (not the peak of the radio emission) at the peak of the MIR images. Assuming the radio jet (PA$=84^{\circ}$) emerges perpendicular to the accretion disk, the accretion disk should be oriented along PA$=-6^{\circ}$.

\begin{figure*}
    \centering
    \includegraphics[width=0.95\linewidth]{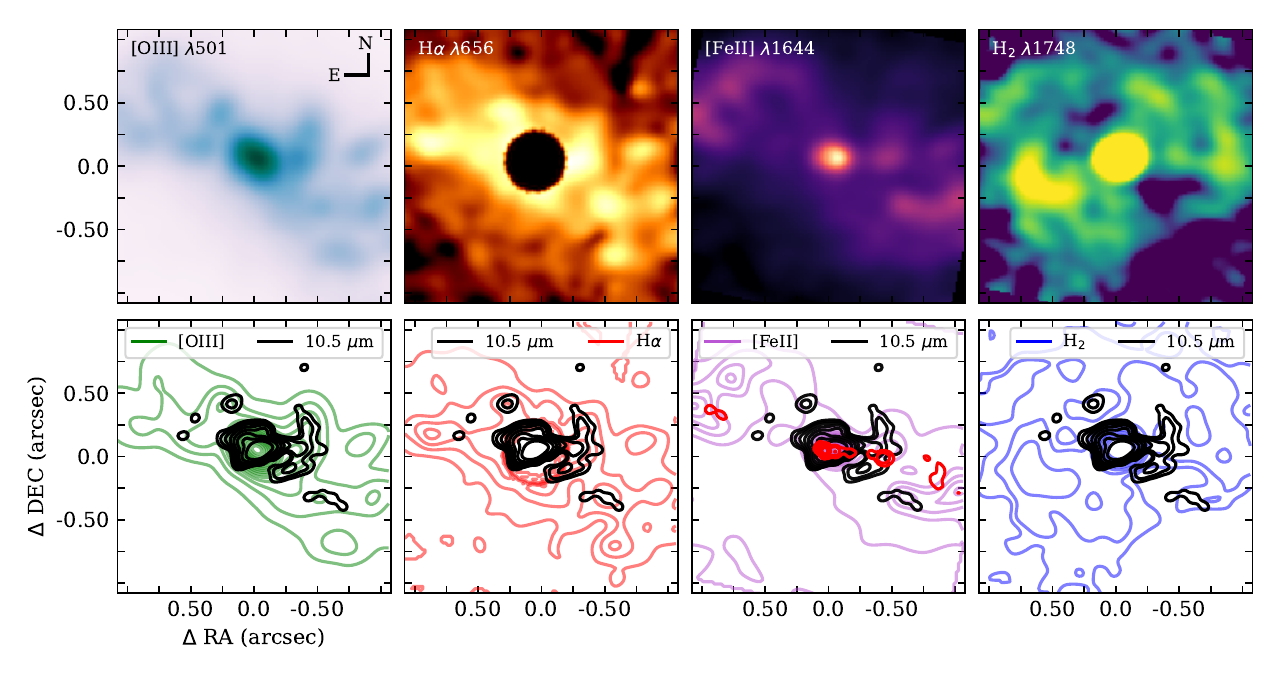}
    \caption{Comparison of MIR and visible/NIR structures.
    The \oiii~image is from \citet{hutchings1999} and the other visible/NIR images are from \cite{may2020}. In the \textit{top row} we show $[\rm{O}{III}]~\lambda501$~nm, H$\alpha~\lambda656$~nm, $[\rm{Fe}{II}]~\lambda1644$~nm, and H$_2~\lambda1748$~nm, from left to right. In the \textit{bottom row} we show contours of the same, with the LBTI 10.5 \micron~contours overplotted (same levels as Fig. \ref{fig:deconv}). In the $[\rm{Fe}{II}]~\lambda1644$~panel, we also show the radio 5 GHz emission from \cite{williams2020}. Many of the MIR features are coincident with visible/NIR emission features. The MIR arc in particular exhibits similarities to the $[\rm{O}III]$ and H$\alpha$ emission. The molecular hydrogen emission is nearly anti-coincident with the visible/MIR features.}
    \label{fig:nir}
\end{figure*}

\subsection{Morphology}
The morphology of NGC 4151 in the four images is consistent in several key ways, but there are also important differences. The $N$-band images (at 8.7 and 10.5 \micron) are consistent with each other. In these images there is a bright central bar extending from northwest to southeast (PA$=115^{\circ}$, 32 pc long). This bar contains the brightest region of each image. Perpendicular to this bar and protruding from the center there are bright extensions to the northeast and southwest, but the extension is brighter to the NE. 

The $LM$-band images (3.7 and 4.8 \micron) are also consistent with each other, but they are considerably less extended than the $N$-band emission. Their smaller extent is not unexpected if one assumes that the $LM$-band emission comes from hotter dust, which should be closer to the accretion disk. Like in the $N$-band images, in the $M$-band image there is a small extension of flux to the NE.

\subsubsection{The Central Bar}
At all wavelengths, there is significant flux along PA$\sim115^{\circ}$, a feature we call the ``central bar.'' The central bar contains the peak flux in each image. To the northeast of the bar, we find extended flux in the 4.8, 8.7, and 10.5 \micron~images. The central bar has a position angle similar to the large scale H$_2~\lambda1748$ torus-like structure \citep[PA$=121^{\circ}$][]{may2020} and with the polarization-inferred torus \citep[PA$=120^{\circ}$][]{ruiz2003}. We therefore assume the emission of the central bar is associated with the so-called ``dusty torus,'' and below we examine three plausible circumnuclear dust models to determine their viability (see \S\ref{sec:physical}). These include (1) a geometrically and optically thick dusty torus; (2) a geometrically thin disk plus wind model; and (3) a wind-only model with no central torus or disk. In all cases, the circumnuclear structure in NGC 4151 should not obscure the BLR or the $\sim 1500$~K dust \citep[which is seen in $K$-band, e.g.,][]{kishimoto2022}.

\subsubsection{The Western Arc}
Additionally, there is a newly observed ``arc'' of emission 0.4" (34 pc) to the west of the center. In both $LM$ images, there is flux extending to the West towards the western $N$-band arc. The extension approximately follows the radio jet (see Fig. \ref{fig:nir}) with PA=$84^{\circ}$. The $LM$-band emission indicates that it is a warm structure starting near the nucleus. 

This large-scale emission was not visible in previous VISIR imaging \citep{asmus2014}, likely due to its lower resolution. To test this we convolved our Fizeau images to the VISIR resolution (see Appendix Fig. \ref{fig:visir}) and were unable to distinguish the feature. On the other hand, previous high-resolution MIDI measurements of the circumnuclear emission exhibited low visibilities \citep{burtscher2013}, due to structures larger than 83 mas (7 pc) being resolved out by the interferometer. The MIDI results could not constrain an orientation of this large structure. 

The western arc's MIR emission is coincident with similar structures observed in \oiii, H$\alpha$, \feii, and [Si VII] (see Fig. \ref{fig:nir}). Despite the fact that the arc is found at the same location as an Airy ring in our imaging, the facts that it (1) is coincident with these other tracers, and (2) is visibly rotating on sky as NGC 4151 transits (see Appendix \ref{app:arc}) give strong arguments for its fidelity. Additionally, its radial extent does not scale with wavelength as one would expect for an Airy ring.  

\citet{may2020} associate the VIS/NIR emission of the western arc with jet-related shocks, and posit that this arc represents a spiral arm of infalling molecular gas that is being impacted by the jet. This represents a channel to feed the AGN, and the NLR bullets (\S\ref{subsec:nlr}) represent a feedback channel. Our observations cannot directly confirm nor refute this hypothesis, but our four-band imaging allows us to measure the temperature of the dust embedded in the structure (see \S\ref{sec:heating}). The observed warm, dusty emission emission extends from the nucleus toward this structure along the radio jet, arguing in favor of outflowing material. However, there may be projection effects that are entangling emission from the in- and outflows, so we leave this as an open question.

\subsubsection{Dusty NLR Clouds}
\label{subsec:nlr}
In the Fizeau images there are a number of faint point sources dispersed around the primary emission. These are likely not imaging artifacts because the majority of these faint sources are (1) consistent between the two Fizeau images, and (2) coincident with VIS/NIR emission. Specifically, the MIR emission is spatially coincident with so-called ``bullets'' tracing a fast outflow. The bullets are observed in Br$10$, H$_2$, \feii, and [Si VII] with velocities $\gtrsim 300$~km~s$^{-1}$  \citep{may2020}. Since the MIR emission is expected to be from the dust continuum -- as our filters are relatively broad and there are no strong emission lines within our filters -- the spatial coincidence with the MIR emission is an indication of dusty outflows. 

SED fitting of a torus + dusty NLR + diffuse dust model to \textit{Spitzer}/IRS 2--35~\micron~spectra by \citet{mor2009} found that the dusty NLR component was crucial. Later work refers to a similar structure as a ``dusty wind'' \citep[e.g., ][]{honig2017, stalevski2019}. Through analysis of 26 type 1 AGN, \citet{mor2009} found that the dusty NLR clouds were typically at a distance of $\sim700$ times the sublimation radius. In NGC 4151, the sublimation radius is 0.03~pc \citep{kishimoto2022}. We measure the clouds at distances 40--80~pc (after deprojecting by 45$^{\circ}$), corresponding to 1300--2600 times the sublimation radius. While these values are larger than the average in \citet{mor2009}, they represent upper limits due to individual clouds rather than a system average, which naturally includes dusty outflows much closer to the nucleus.

\subsection{Heating and Emission Mechanisms}
\label{sec:heating}
As expected, the material closest to the accretion disk exhibits the highest temperatures. We measure $\sim 500$~K in this central region. It is unsurprising that this is much cooler than the 1200--1500 K sublimation region measured in the $K$-band due to our large pixel size. The hot $K$-band dust was well fit by a thin ring with radius 0.5 mas \citep[0.03 pc;][]{kishimoto2022}, and so the hot dust emission fills only a minuscule fraction of the emitting surface area within an $18\times18$~mas pixel. It is largely drowned out by the more extended, cooler dust when averaging over the surface area within the pixel.

Heating from the AGN alone cannot explain the observed flux profile and inferred temperature profile. Dust heated by the AGN should follow the temperature profile of \citet{barvainis1987}, which falls radially $T(r) \propto r^{-2/5.6}$. Not only is the region around the nucleus not azimuthimally symmetric (with increased temperature in a narrow line toward the NE), but there is also an increase in temperature $\sim 0.4$" from the nucleus within the western arc. 

A high temperature region (relative to temperatures at similar radii) extends from the nucleus along PA$\approx 30^{\circ}$. This PA follows the northern wall of the ionization cone. 
Notably, the hot sublimation zone also shows an extension along PA$=19\pm10^{\circ}$ \citep{kishimoto2022}, hinting at a possible hot outflow.

\begin{figure*}[t]
    \centering
    \includegraphics[width=0.99\linewidth]{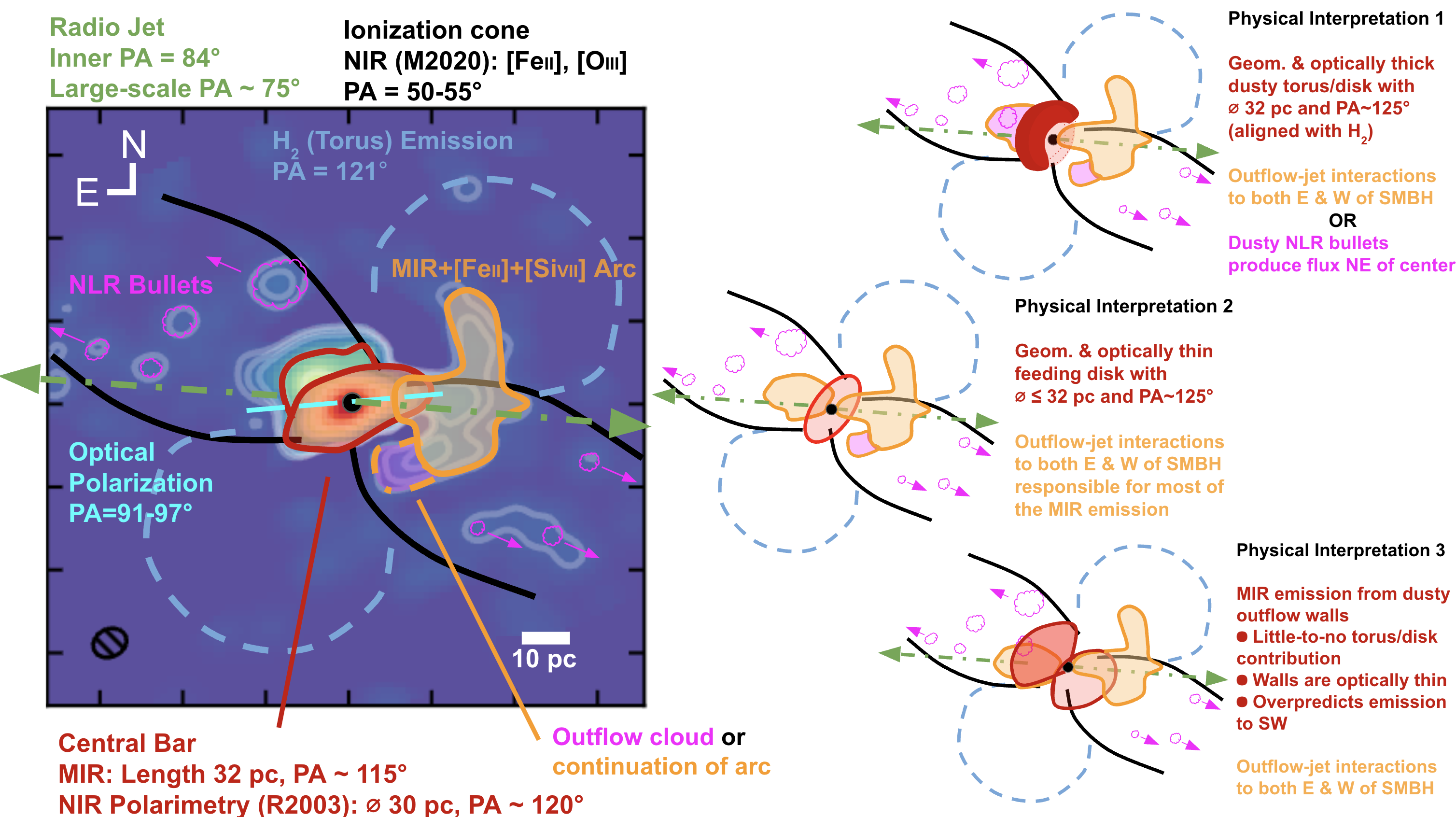}
    \caption{Summary sketch. The NIR components are based on \citet{may2020}, the NIR polarimetry results are from \citet{ruiz2003}, and the radio jet orientation is from \citet{williams2020}. The ionization cone and circumnuclear dust structures are misaligned with the radio jet/accretion disk. The inner dot-dashed, cyan line represents the PA of the optical polarization \citep[PA$=91-97^{\circ}$][]{martel1998}, which is even farther misaligned from the ionization cone. We present three physical interpretations which aim to explain the observed morphology. We favor Physical Interpretation 2 for reasons described in \S\ref{sec:physical}. Note that the western arc's southernmost region is due to either outflowing material or a continuation of the shock front, depending on the physical model. 
    }
    \label{fig:sketch}
\end{figure*}

\subsubsection{Indications of a Shock Front}
In both $LM$ images, there is flux extending to the West towards the western $N$-band arc. The radio emission near the SMBH also shows that the jet leaves the nucleus to the West, coincident with the MIR features. The $LM$-band fluxes indicate that warm dust is co-spatial with the jet as it leaves the nucleus. This is consistent with the measurements of NGC 1068 by \citet{isbell2025}, which indicated enhanced MIR fluxes along the jet direction. Our interpretation was that heating from radio jet shocks was responsible for the excess in MIR emission. A similar interpretation is plausible in NGC 4151.

In the western arc we find that a radio knot, an ionized cloud (showing H$\alpha$, \feii, \oiii, and [SiVI]~emission), and the MIR emission are all co-spatial (in projection). 
At this same location, we measure a significant increase in temperature and silicate feature depth. Given the region's coincidence with the ionized cloud and proximity to a knot of the radio jet, it is very likely that the region is being shock heated by the radio jet interacting with the ISM. Moreover, this local increase in temperature is far above what is expected from heating by the accretion disk \citep[which should follow $T(r) \propto r^{-2/5.6}$;][]{barvainis1987}.

Additionally, we measure a deeper silicate absorption feature in this cloud than elsewhere along the jet or toward the nucleus. The higher fitted A$_{\rm V}$ values indicate either that the MIR emission is more heavily obscured by dust and/or that there is a higher concentration of silicates in this arc. There is unlikely to be more galaxy-scale foreground absorption at only this line of sight, so this effect seems local to the western arc. It most likely indicates the three-dimensional nature of the structure. The emitting clouds could be embedded within a larger structure that is being locally impacted by the jet. It could also be that the expanding shock front is expanding toward the observer, increasing the line-of-sight absorption. The latter idea is less likely, since the measured velocities of the \feii~in this region are close to systemic \citep{may2020}. Alternatively, the composition of the dust (i.e., the relative amounts of carbon and silicates) can give insights into the dust's history \citep[e.g.,][]{schartmann2005, tsuchikawa2021}. For example, since the sublimation temperatures of silicate and graphite dust are different \citep[1000~K vs 1500~K, respectively][]{schartmann2005}, dust that at one time was near the sublimation zone might be depleted in silicates. As we see enhanced silicate absorption in the western arc, it would be unlikely that the dust there was previously near the nucleus, arguing against this structure being related to an outflow. Due to local changes in dust composition, our adoption of a global $\kappa_{\lambda}$ in the modified blackbody fitting is likely overly simplistic. Either spatially resolved spectral information (from e.g., JWST or LBTI/ALES) or LBTI images in more filters would be necessary to further constrain the dust composition distribution in this source.

\subsection{Physical Interpretations}
\label{sec:physical}
In this section we consider three physical interpretations of the observed structures. Each interpretation is based on plausible results from previous SED fitting, radiative transfer modeling, or high-resolution observations of other nearby AGN. Each of the interpretations is sketched in Fig. \ref{fig:sketch}. 

In general we note that it is difficult to justify the claim that the central bar ($PA\sim115^{\circ}$) directly represents a disk or torus. None of the physical models we consider easily support a torus/disk misaligned with \textit{both} the larger scale molecular structures and the accretion disk. Instead, in each considered model the observed PA of the central bar is due to a blending of the emission from the disk/torus/wind (PA$=120-125^{\circ}$) and the shock-heated material along the radio jet (PA$=84^{\circ}$). Finally, in each model we assume an inclination of 40--45$^{\circ}$ away from edge-on \citep{kishimoto2022}.

\subsubsection{The Classical Dusty Torus}
The first model is a geometrically and optically thick torus with a diameter of $\sim 32$~pc and a PA$=125^{\circ}$. In this case, the MIR emission is primarily due re-radiated thermal emission along the surface of the torus, which is heated by the accretion disk. Near the peak flux of our images, this model predicts that we should see the far side of the torus as well as the sublimation zone and BLR. The extension of flux we observe toward the NE of the nucleus could either be due to outflow-jet interactions (similar to what we see to the West) or lucky placement of dusty NLR ``bullets'' which we see farther out. 

An optically thick torus should block the emission of visible and NIR lines. However, as is shown in Fig. \ref{fig:nir}, there is ample \oiii and \feii emission at the same location as the inferred torus. Since these lines are associated with ionization from direct illumination of the accretion disk, they are extremely unlikely to be found along the surface of a thermally emitting torus. It is therefore unlikely that the circumnuclear structure in NGC 4151 is optically thick. A clumpy torus with a low dust density distribution \cite[see e.g., the models in][]{nenkova2008, hoenig2010, stalevski2016} could better match the observations, and it would behave similarly to ``The Disk+Wind Model'' described below.

\subsubsection{The Disk+Wind Model}
The second model is a geometrically and optically thin disk with a diameter of $\sim 32$~pc and a PA$=125^{\circ}$. We assume symmetric outflow cones above and below the disk. In this model, the MIR emission along $PA=125^{\circ}$ is also re-radiated thermal emission, but it is blended with the shock heated emission from outflowing material along the radio jets to the E and W. Note here that the outflows are associated with the radio jet rather than the typical radiation-driven fountains (see ``The Wind-Only Model'' below for our reasoning). 

The disk is assumed to be optically thin at essentially all radii as one looks through it (to match Sy2 observations, the integrated optical depth along the disk will necessarily be high). This allows for the visibility of visible and NIR emission line features found in the NLR. The optical thinness of the disk also gives a more direct view to the outflow to the E of the nucleus, which is blended with the disk's emission.

\subsubsection{The Wind-Only Model}
Lastly we consider a model in which the disk is extremely small or nonexistent, and the radiation-pressure driven wind is the dominant MIR emission source. The winds have the same opening angle as the ionization cones, and we assume symmetry to either side of the accretion disk. The walls of the wind's outflow cone are optically thin \citep[due to clumpiness, e.g., ][]{honig2017,stalevski2019}. 

To the NE of the nucleus, the blending of the dusty wind and the outflow along the jet can morphologically match the observed emission. A symmetric cone to the SW, however, would greatly overpredict the emission when compared to our observations. Moreover, the flux emitted by such an outflow cone is shown via RT modeling to drop rapidly with distance from the nucleus. In NGC 1068 -- a much more luminous AGN -- the outflow cone already became too faint to detect beyond $\sim 15$~pc \citep{isbell2025}. In the less luminous NGC 4151, the cone would become too faint at a much smaller distance. Based neither on simple morphological comparisons nor on physical reasoning is the wind-only interpretation likely. If there are radiation-driven fountain flows in NGC 4151, they are on much smaller scales than would be necessary to produce all of the observed structures.

\subsubsection{Comparisons}
While none of these relatively simple models match the observations perfectly, we prefer the disk+wind model (Physical Interpretation 2 in Fig. \ref{fig:sketch}) for a number of reasons given below.

\textbf{The obscuring structure must be optically thin.} Our imaging reveals individual dusty NLR clouds and emission associated with the edge of the ionization cone. The latter emission is \textit{brighter} to the NE than equivalent emission to the SW. Present also are bright visible and NIR emission lines that an optically thick structure would obscure (see Fig. \ref{fig:nir}). Therefore, the disk or torus in NGC 4151 must be clumpy, optically thin, or very compact. 

The analogous structure imaged in NGC 1068 causes significant extinction far to the south of the nucleus \citep{gamezrosas2022, isbell2025}, quite unlike what we observe here. Our imaged disk or torus appears to be relatively compact. The emission along PA $\approx 125^{\circ}$ is measured to be $\leq 32$~pc. Using UKIRT polarimetry, \citet{ruiz2003} inferred a scattering structure (i.e., torus) with diameter 24 pc at PA $120^{\circ}$. This result is consistent with the size and orientation of our inferred structures. However, it is important to note that NIR polarization does not necessarily trace a dusty torus. Instead, it may trace an equatorial scattering region surrounding the accretion disk \citep[see ][and the references therein]{kishimoto2008}. This interpretation could indicate that the \citet{ruiz2003} structures trace a geometrically thin disk rather than a torus, which would further support our disk+wind interpretation.

\textbf{The circumnuclear dust should trace inflowing as well as outflowing material.} Hydrodynamical modeling shows that a thin equatorial molecular disk likely supplies material to the SMBH, while fountain-like outflows -- caused by radiation pressure -- transport material away \citep[e.g.,][]{wada2016}. In NGC 4151, the molecular hydrogen (H$_{2}~\lambda1748$) distribution reported by \citet{may2020} is primarily found in the ``shadow of the ionization cones'' and has PA$=121^{\circ}$. This puts it in excellent agreement with the torus/disk as inferred from our images and via UKIRT polarimetry. Additionally, \citet{may2020} hypothesize that the PA$=121^{\circ}$ H$_2$ emission could be associated with inflowing material which feeds the AGN. High velocity red/blue shifts in the H$_2$ emission are measured within the ionization cone, and are associated with outflows. It appears that the H$_2$ emission traces more than one structure, so its overall morphology becomes hard to interpret. Nonetheless, inflowing H$_2$ emission (PA$=121^{\circ}$) would fit the disk+wind flavor of the circumnuclear dust\citep[e.g.,][]{honig2017,stalevski2019} wherein the disk/torus contains the material that feeds the AGN, and the dusty wind traces outflows. The wind-only model is unlikely because it would be transient, as the accretion disk would be starved of material with which to drive the active phase.

\textbf{There should be little-to-no emission to the SW of the nucleus.} This is not a physical justification, but an observational one. In none of our imaged wavelengths do we find significant emission extending from the nucleus toward the SW. This argues against the wind-only model, which should produce symmetric emission to the NE and SW. We note, however, that this also limits the scale of the cone-line outflow predicted by the disk+wind models \citep{wada2016, stalevski2019}. While we observe dusty NLR clouds within the ionization cone, it appears that most of the outflow-related emission is associated with the jet rather than a radiation-pressure-driven wind. The bolometric luminosity and Eddington ratio in NGC 4151 are both relatively low, and this may explain the lack of a large-scale radiation-driven fountain -- less material will reach escape velocities and will instead fall back toward the disk/torus.

Neither this nor previous publications support a disk/torus that aligns well with the radio jet or accretion disk rotation axis. Typically, the inner jet orientation is taken to be the rotation axis of the accretion disk (AD) and the torus, meaning that edges of the AD and the disk/torus should be perpendicular to the jet. Importantly, the inner jet orientation is not necessarily the same as the large-scale orientation: at large scales, the jet has PA$_{\rm jet,~lg} \sim 75^{\circ}$ due to either an inner deflection or precession \citep{ulvestad1998}. In NGC 1068, for example, the jet is apparently deflected by a hot, dusty cloud $\sim 25$~pc from the nucleus \citep{may2017,isbell2025}.
In NGC 4151 because the jet has an inner PA$_{\rm jet} = 84^{\circ}$ \citep{mundell2003}, the AD and torus should have PA$_{\rm torus,~exp} \in [167, 180]^{\circ}$; it is measured rather to have PA$_{\rm torus,~obs} = 125^{\circ}$. Due to the large offset between the jet and the dusty, molecular torus, the jet could be impacting the cool, inflowing material. This material would then be shock-heated; we see ample evidence in NGC 4151 of such shock heating (\S\ref{sec:heating}). Such an interaction could also explain the change in jet orientation at different scales in NGC 4151 via deflection off the inflowing material.

\subsection{Unified Model of AGN}
\label{sec:unified-model}

The physical parameters of NGC 4151 and NGC 1068 are given in Table \ref{tab:comparison}. While the distances, infrared luminosities, and black hole masses of the two objects are similar, the bolometric luminosity and Eddington ratio of NGC 1068 are much larger than those of NGC 4151. The inner inclinations (i.e., measured in the central $\leq 100$~pc) are also quite different, with NGC 1068 approximately edge-on and NGC 4151 inclined $45^{\circ}$. 

The difference in bolometric luminosity could explain the difference in measured disk diameters. A dust size-luminosity relationship of the form $r \propto L^{0.5}$ has been observed in a large number of AGN in the NIR \citep[e.g.,][]{kishimoto2011,koshida2014, li2023, gravity2024}. This trend  may be a bit flatter in the MIR, but it was a good approximation for the 26 MIDI-observed AGN \citep{burtscher2013}.
Relating the bolometric luminosities of NGC 4151 and NGC 1068 to the measured NGC 4151 disk diameter of 32 pc, we would expect a disk of diameter 76 pc in NGC 1068 (using the $r\propto L^{0.5}$ scaling). Observationally, \citet{isbell2025} inferred a diameter of $\geq 70$~pc from dust obscuration in NGC 1068. These two LBTI-observed objects are consistent with expected scaling relations in the MIR. 

Overall, the observed morphologies of the two sources are quite similar when inclination effects are taken into account. They consist of:
\begin{itemize}
\item A dense (obscuring) dusty structure approximately perpendicular to the ionization cone. In NGC 4151 this is seen in MIR emission, while in NGC 1068 it is seen in absorption. 
\item Extended MIR emission within the ionization cone and edge-brightening effects at the edges of the cone. In NGC 4151 there is additional MIR emission associated with the outflowing ``bullets'' of \citet{may2020}. This extended emission is sometimes called the ``polar wind'' and is attributed to radiation-pressure driven outflows \citep[e.g., ][]{wada2016}.  
\item Bright, warm dust clouds associated with interactions between the radio jet and the ISM. In NGC 4151 we measure a local increase in temperature in this region, likely explained by shock heating. In both AGN, the inner PA of the radio jet differs from the large-scale jet PA, indicating deflection off of nuclear clouds, jet precession, or both. 
\end{itemize}

Notably, the nuclear jet-ISM interaction represents a large fraction of the MIR flux. In RT modeling of AGN dust emission (e.g., the disk+wind model), all emission is thought to arise from heating by the accretion disk. This is based on SED fitting to fluxes from images which do not resolve any of the nuclear substructures \citep[e.g., WISE with resolution 6";][]{wright2010}. In both AGN imaged with the LBTI, however, secondary heating from the radio jet appears to play a large role. Additionally, recent JWST results (at $\sim2$~pc resolution) in the Sy2 Circinus Galaxy indicate significant amounts of jet-heated dust rather than purely AD-heated dust \citep{lopezrodriguez2025}. Very few AGN tori can be resolved directly, and instead SED fitting or color analysis has historically given morphological constraints on the torus \citep[e.g.,][]{honig2017}. Without taking jet heating into account, however, previous RT models (driven by SED fitting) may have misattributed flux to certain structures and biased the morphological results. Despite the need to extend the RT modeling to include shock heating, we find excellent qualitative agreement between our observations and the disk+wind version of the Unified Model. 

\begin{table}[t]
\centering
\begin{tabular}{l|cc}
\hline
\textbf{Property} & \textbf{NGC 4151} & \textbf{NGC 1068} \\
\hline
Type & Sy1/Sy1.5 & Sy2 \\
Distance [Mpc] & 15.9–19.9$^a$ & 14.4$^b$ \\
$\log_{10}(L_{12\mu\rm{m}} / \rm{W~Hz}^{-1}$) & 36.01$^c$ & 36.9$^c$ \\
$\log_{10}(L_{\rm bol} / \rm{erg~s}^{-1})$ & $43.8^d$ & 44.6–45.7$^e$ \\
$\log_{10}(M_{\rm SMBH} / M_\odot$) & $7.7^f$ & 6.9–7.2$^e$\\
Eddington Ratio & 0.01–0.1$^g$ & 0.2–1+$^{e,*}$ \\
\hline
Inner Inclination [$^\circ$] & 40-45$^h$ & 75$^i$ \\
Torus Diameter [pc] & 32 & 70+$^j$ \\
\hline
\end{tabular}
\caption{Comparison of key properties between NGC 4151 and NGC 1068.
$a$: \cite{yuan2020, honig2014}, 
$b$: \cite{blandhawthorn1997},
$c$: \cite{leftley2019}, 
$d$: \cite{kaspi2005}, 
$e$: \cite{gravity2020},
$f$: \cite{honig2014},
$g$: \cite{merritt2022}, 
$h$: \cite{kishimoto2022}, 
$i$:\cite{leftley2024}, 
$j$: \cite{isbell2025}. *Some authors \citep[e.g.,][]{gravity2020, leftley2019} calculate that NGC 1068 is emitting at or above the Eddington limit. 
}
\label{tab:comparison}
\end{table}

\section{Conclusions}
\label{sec:conc}
In this paper, we present mid-infrared (MIR) imaging of NGC 4151 using the Large Binocular Telescope Interferometer (LBTI). In the $N$-band, we employed the open-loop Fizeau interferometric mode of the LBTI to obtain 66–104 mas resolution, effectively leveraging the full 28.8 m baseline of the LBTI. Additional AO-assisted imaging in the $LM$-bands achieved angular resolutions of 50–62 mas using one of the LBT's 8.4 m apertures. Our images offer spatial resolution intermediate between existing VLTI/MIDI and VLT/VISIR observations, providing a preview of the capabilities expected from future Extremely Large Telescopes (ELTs) in the MIR.

Our MIR images directly resolve a nuclear structure, likely corresponding to the so-called dust torus, with a diameter of 32 pc and position angle (PA) of $125^{\circ}$. This is consistent with prior UKIRT polarimetry \citep[24 pc, PA$=120^{\circ}$;][]{ruiz2003} and with a putative torus of molecular hydrogen \citep[PA$=121^{\circ}$;][]{may2020}. Notably, this orientation deviates significantly from the PA perpendicular to the radio jet ($\sim0^{\circ}$), challenging the standard assumption of torus orientation. Whether we interpret this structure as a torus or a geometrically thin disk, we find that it must be optically thin in order to allow observed visible and NIR emission lines to pass through.

By convolving all four images to a common resolution, we extracted spatially resolved spectral energy distributions (SEDs) across the nuclear region. Modified blackbody fits to these SEDs allowed us to map dust color temperature and extinction profiles. These temperature distributions reveal contributions from both the central engine and jet-related shock heating. Spatial alignment between MIR emission, optical/NIR line emission (\oiii, \feii, H$\alpha$), and radio structures further supports the shock heating scenario.

A comparative analysis with LBTI images of the Seyfert 2 galaxy NGC 1068 provided a valuable test of the Unified Model of AGN. Despite differences in Eddington ratio, both AGN exhibit similar MIR structures. A composite model featuring a geometrically thin disk/torus and a dusty wind (``disk+wind") appears consistent with both sources. The torus in NGC 4151 is notably more compact than in NGC 1068, in line with the expected $r \propto L^{0.5}$ scaling. In both cases, we observe evidence of shock heating from radio jets on scales comparable to the dusty torus.

These LBTI observations of two nearby AGN highlight a potentially missing component in current models of AGN circumnuclear structure: shock heating by the radio jet. This process appears to influence the thermal state of dust out to scales of $\sim$100 pc, yet it is not accounted for in most radiative transfer models, which typically assume dust heating is dominated by radiation from the accretion disk. To refine our understanding of AGN environments, further high-resolution observations are needed to assess the prevalence of this disk+wind+shock configuration. Moreover, detailed radiation-hydrodynamical simulations will be essential to model the interplay between radiative and mechanical feedback in shaping the circumnuclear regions of AGN. Accounting for shock heating is critical, as current SED-based interpretations of AGN structure may be systematically biased by neglecting this additional energy source.

\section*{~}

\textit{Acknowledgments}: The LBT is an international collaboration among institutions in the United States, Italy, and Germany. LBT Corporation Members are: The University of Arizona on behalf of the Arizona Board of Regents; Istituto Nazionale di Astrofisica, Italy; LBT Beteiligungsgesellschaft, Germany, representing the Max-Planck Society, The Leibniz Institute for Astrophysics Potsdam, and Heidelberg University; The Ohio State University (OSU), representing OSU, University of Notre Dame, University of Minnesota and University of Virginia. Observations have benefited from the use of ALTA Center (alta.arcetri.inaf.it) forecasts performed with the Astro-Meso-Nh model. Initialization data of the ALTA automatic forecast system come from the General Circulation Model (HRES) of the European Centre for Medium Range Weather Forecasts.

MK acknowledges the support from the JSPS grant \#24K00679. The authors thank Thomas Stuber for insightful discussions during the preparation of this work. The authors also thank the anonymous referee for their helpful comments.


%

\vspace{5mm}
\facility{Large Binocular Telescope}


\software{astropy \citep{astropy2022}, CASA \citep{casa2022}, scikit-image \citep{scikit-image}, numpy \citep{numpy2020}, scipy \citep{scipy2020}, matplotlib \citep{matplotlib}, LIZARD (\href{https://github.com/jwisbell/lbti_fizeau}{https://github.com/jwisbell/lbti\_fizeau})}



\bibliography{main}{}

@ARTICLE{kishimoto2022,
       author = {{Kishimoto}, Makoto and {Anderson}, Matthew and {ten Brummelaar}, Theo and {Farrington}, Christopher and {Antonucci}, Robert and {H{\"o}nig}, Sebastian and {Millour}, Florentin and {Tristram}, Konrad R.~W. and {Weigelt}, Gerd and {Sturmann}, Laszlo and {Sturmann}, Judit and {Schaefer}, Gail and {Scott}, Nic},
        title = "{The Dust Sublimation Region of the Type 1 AGN NGC 4151 at a Hundred Microarcsecond Scale as Resolved by the CHARA Array Interferometer}",
      journal = {\apj},
         year = 2022,
        month = nov,
       volume = {940},
       number = {1},
          eid = {28},
        pages = {28},
          doi = {10.3847/1538-4357/ac91c4},
archivePrefix = {arXiv},
       eprint = {2209.06061},
 primaryClass = {astro-ph.GA},
       adsurl = {https://ui.adsabs.harvard.edu/abs/2022ApJ...940...28K}
}

@ARTICLE{isbell2025,
       author = {{Isbell}, J.~W. and {Ertel}, S. and {Pott}, J. -U. and {Weigelt}, G. and {Stalevski}, M. and {Leftley}, J. and {Jaffe}, W. and {Petrov}, R.~G. and {Moszczynski}, N. and {Vermot}, P. and {Hinz}, P. and {Burtscher}, L. and {G{\'a}mez Rosas}, V. and {Becker}, A. and {Carlson}, J. and {Faramaz-Gorka}, V. and {Hoffmann}, W.~F. and {Leisenring}, J. and {Power}, J. and {Wagner}, K.},
        title = "{Direct imaging of active galactic nucleus outflows and their origin with the 23 m Large Binocular Telescope}",
      journal = {Nature Astronomy},
         year = 2025,
        month = mar,
       volume = {9},
        pages = {417-427},
          doi = {10.1038/s41550-024-02461-y},
archivePrefix = {arXiv},
       eprint = {2502.01840},
 primaryClass = {astro-ph.GA},
       adsurl = {https://ui.adsabs.harvard.edu/abs/2025NatAs...9..417I}
}

@ARTICLE{nenkova2008,
       author = {{Nenkova}, Maia and {Sirocky}, Matthew M. and {Nikutta}, Robert and {Ivezi{\'c}}, {\v{Z}}eljko and {Elitzur}, Moshe},
        title = "{AGN Dusty Tori. II. Observational Implications of Clumpiness}",
      journal = {\apj},
         year = 2008,
        month = sep,
       volume = {685},
       number = {1},
        pages = {160-180},
          doi = {10.1086/590483},
archivePrefix = {arXiv},
       eprint = {0806.0512},
 primaryClass = {astro-ph},
       adsurl = {https://ui.adsabs.harvard.edu/abs/2008ApJ...685..160N}
}

@ARTICLE{tristram2014,
       author = {{Tristram}, Konrad R.~W. and {Burtscher}, Leonard and {Jaffe}, Walter and {Meisenheimer}, Klaus and {H{\"o}nig}, Sebastian F. and {Kishimoto}, Makoto and {Schartmann}, Marc and {Weigelt}, Gerd},
        title = "{The dusty torus in the Circinus galaxy: a dense disk and the torus funnel}",
      journal = {\aap},
         year = 2014,
        month = mar,
       volume = {563},
          eid = {A82},
        pages = {A82},
          doi = {10.1051/0004-6361/201322698},
archivePrefix = {arXiv},
       eprint = {1312.4534},
 primaryClass = {astro-ph.GA},
       adsurl = {https://ui.adsabs.harvard.edu/abs/2014A&A...563A..82T}
}

@ARTICLE{isbell2022,
       author = {{Isbell}, J.~W. and {Meisenheimer}, K. and {Pott}, J. -U. and {Stalevski}, M. and {Tristram}, K.~R.~W. and {Sanchez-Bermudez}, J. and {Hofmann}, K. -H. and {G{\'a}mez Rosas}, V. and {Jaffe}, W. and {Burtscher}, L. and {Leftley}, J. and {Petrov}, R. and {Lopez}, B. and {Henning}, T. and {Weigelt}, G. and {Allouche}, F. and {Berio}, P. and {Bettonvil}, F. and {Cruzalebes}, P. and {Dominik}, C. and {Heininger}, M. and {Hogerheijde}, M. and {Lagarde}, S. and {Lehmitz}, M. and {Matter}, A. and {Meilland}, A. and {Millour}, F. and {Robbe-Dubois}, S. and {Schertl}, D. and {van Boekel}, R. and {Varga}, J. and {Woillez}, J.},
        title = "{The dusty heart of Circinus. I. Imaging the circumnuclear dust in N-band}",
      journal = {\aap},
         year = 2022,
        month = jul,
       volume = {663},
          eid = {A35},
        pages = {A35},
          doi = {10.1051/0004-6361/202243271},
archivePrefix = {arXiv},
       eprint = {2205.01575},
 primaryClass = {astro-ph.GA},
       adsurl = {https://ui.adsabs.harvard.edu/abs/2022A&A...663A..35I}
}

@ARTICLE{isbell2023,
       author = {{Isbell}, J.~W. and {Pott}, J. -U. and {Meisenheimer}, K. and {Stalevski}, M. and {Tristram}, K.~R.~W. and {Leftley}, J. and {Asmus}, D. and {Weigelt}, G. and {G{\'a}mez Rosas}, V. and {Petrov}, R. and {Jaffe}, W. and {Hofmann}, K. -H. and {Henning}, T. and {Lopez}, B.},
        title = "{The dusty heart of Circinus. II. Scrutinizing the LM-band dust morphology using MATISSE}",
      journal = {\aap},
         year = 2023,
        month = oct,
       volume = {678},
          eid = {A136},
        pages = {A136},
          doi = {10.1051/0004-6361/202347307},
archivePrefix = {arXiv},
       eprint = {2309.07613},
 primaryClass = {astro-ph.GA},
       adsurl = {https://ui.adsabs.harvard.edu/abs/2023A&A...678A.136I}
}

@ARTICLE{gamezrosas2022,
       author = {{G{\'a}mez Rosas}, Violeta and {Isbell}, Jacob W. and {Jaffe}, Walter and {Petrov}, Romain G. and {Leftley}, James H. and {Hofmann}, Karl-Heinz and {Millour}, Florentin and {Burtscher}, Leonard and {Meisenheimer}, Klaus and {Meilland}, Anthony and {Waters}, Laurens B.~F.~M. and {Lopez}, Bruno and {Lagarde}, St{\'e}phane and {Weigelt}, Gerd and {Berio}, Philippe and {Allouche}, Fatme and {Robbe-Dubois}, Sylvie and {Cruzal{\`e}bes}, Pierre and {Bettonvil}, Felix and {Henning}, Thomas and {Augereau}, Jean-Charles and {Antonelli}, Pierre and {Beckmann}, Udo and {van Boekel}, Roy and {Bendjoya}, Philippe and {Danchi}, William C. and {Dominik}, Carsten and {Drevon}, Julien and {Gallimore}, Jack F. and {Graser}, Uwe and {Heininger}, Matthias and {Hocd{\'e}}, Vincent and {Hogerheijde}, Michiel and {Hron}, Josef and {Impellizzeri}, Caterina M.~V. and {Klarmann}, Lucia and {Kokoulina}, Elena and {Labadie}, Lucas and {Lehmitz}, Michael and {Matter}, Alexis and {Paladini}, Claudia and {Pantin}, Eric and {Pott}, J{\"o}rg-Uwe and {Schertl}, Dieter and {Soulain}, Anthony and {Stee}, Philippe and {Tristram}, Konrad and {Varga}, Jozsef and {Woillez}, Julien and {Wolf}, Sebastian and {Yoffe}, Gideon and {Zins}, Gerard},
        title = "{Thermal imaging of dust hiding the black hole in NGC 1068}",
      journal = {\nat},
         year = 2022,
        month = feb,
       volume = {602},
       number = {7897},
        pages = {403-407},
          doi = {10.1038/s41586-021-04311-7},
archivePrefix = {arXiv},
       eprint = {2112.13694},
 primaryClass = {astro-ph.GA},
       adsurl = {https://ui.adsabs.harvard.edu/abs/2022Natur.602..403G}
}

@ARTICLE{stalevski2019,
       author = {{Stalevski}, Marko and {Tristram}, Konrad R.~W. and {Asmus}, Daniel},
        title = "{Dissecting the active galactic nucleus in Circinus - II. A thin dusty disc and a polar outflow on parsec scales}",
      journal = {\mnras},
         year = 2019,
        month = apr,
       volume = {484},
       number = {3},
        pages = {3334-3355},
          doi = {10.1093/mnras/stz220},
archivePrefix = {arXiv},
       eprint = {1901.05488},
 primaryClass = {astro-ph.GA},
       adsurl = {https://ui.adsabs.harvard.edu/abs/2019MNRAS.484.3334S}
}

@ARTICLE{garcia-bernete2022,
       author = {{Garc{\'\i}a-Bernete}, I. and {Gonz{\'a}lez-Mart{\'\i}n}, O. and {Ramos Almeida}, C. and {Alonso-Herrero}, A. and {Mart{\'\i}nez-Paredes}, M. and {Ward}, M.~J. and {Roche}, P.~F. and {Acosta-Pulido}, J.~A. and {L{\'o}pez-Rodr{\'\i}guez}, E. and {Rigopoulou}, D. and {Esparza-Arredondo}, D.},
        title = "{Torus and polar dust dependence on active galactic nucleus properties}",
      journal = {\aap},
         year = 2022,
        month = nov,
       volume = {667},
          eid = {A140},
        pages = {A140},
          doi = {10.1051/0004-6361/202244230},
archivePrefix = {arXiv},
       eprint = {2210.03508},
 primaryClass = {astro-ph.GA},
       adsurl = {https://ui.adsabs.harvard.edu/abs/2022A&A...667A.140G}
}

@ARTICLE{wada2016,
       author = {{Wada}, Keiichi and {Schartmann}, Marc and {Meijerink}, Rowin},
        title = "{Multi-phase Nature of a Radiation-driven Fountain with Nuclear Starburst in a Low-mass Active Galactic Nucleus}",
      journal = {\apjl},
         year = 2016,
        month = sep,
       volume = {828},
       number = {2},
          eid = {L19},
        pages = {L19},
          doi = {10.3847/2041-8205/828/2/L19},
archivePrefix = {arXiv},
       eprint = {1608.06995},
 primaryClass = {astro-ph.GA},
       adsurl = {https://ui.adsabs.harvard.edu/abs/2016ApJ...828L..19W}
}

@ARTICLE{williamson2020,
       author = {{Williamson}, David and {H{\"o}nig}, Sebastian and {Venanzi}, Marta},
        title = "{Radiation Hydrodynamics Models of Active Galactic Nuclei: Beyond the Central Parsec}",
      journal = {\apj},
         year = 2020,
        month = jul,
       volume = {897},
       number = {1},
          eid = {26},
        pages = {26},
          doi = {10.3847/1538-4357/ab989e},
archivePrefix = {arXiv},
       eprint = {2006.00918},
 primaryClass = {astro-ph.GA},
       adsurl = {https://ui.adsabs.harvard.edu/abs/2020ApJ...897...26W}
}

@ARTICLE{honig2014,
       author = {{H{\"o}nig}, Sebastian F. and {Watson}, Darach and {Kishimoto}, Makoto and {Hjorth}, Jens},
        title = "{A dust-parallax distance of 19 megaparsecs to the supermassive black hole in NGC 4151}",
      journal = {\nat},
         year = 2014,
        month = nov,
       volume = {515},
       number = {7528},
        pages = {528-530},
          doi = {10.1038/nature13914},
archivePrefix = {arXiv},
       eprint = {1411.7032},
 primaryClass = {astro-ph.GA},
       adsurl = {https://ui.adsabs.harvard.edu/abs/2014Natur.515..528H}
}

@ARTICLE{gravity2020,
       author = {{GRAVITY Collaboration} and {Pfuhl}, O. and {Davies}, R. and {Dexter}, J. and {Netzer}, H. and {H{\"o}nig}, S. and {Lutz}, D. and {Schartmann}, M. and {Sturm}, E. and {Amorim}, A. and {Brandner}, W. and {Cl{\'e}net}, Y. and {de Zeeuw}, P.~T. and {Eckart}, A. and {Eisenhauer}, F. and {F{\"o}rster Schreiber}, N.~M. and {Gao}, F. and {Garcia}, P.~J.~V. and {Genzel}, R. and {Gillessen}, S. and {Gratadour}, D. and {Kishimoto}, M. and {Lacour}, S. and {Millour}, F. and {Ott}, T. and {Paumard}, T. and {Perraut}, K. and {Perrin}, G. and {Peterson}, B.~M. and {Petrucci}, P.~O. and {Prieto}, M.~A. and {Rouan}, D. and {Shangguan}, J. and {Shimizu}, T. and {Sternberg}, A. and {Straub}, O. and {Straubmeier}, C. and {Tacconi}, L.~J. and {Tristram}, K.~R.~W. and {Vermot}, P. and {Waisberg}, I. and {Widmann}, F. and {Woillez}, J.},
        title = "{An image of the dust sublimation region in the nucleus of NGC 1068}",
      journal = {\aap},
         year = 2020,
        month = feb,
       volume = {634},
          eid = {A1},
        pages = {A1},
          doi = {10.1051/0004-6361/201936255},
archivePrefix = {arXiv},
       eprint = {1912.01361},
 primaryClass = {astro-ph.GA},
       adsurl = {https://ui.adsabs.harvard.edu/abs/2020A&A...634A...1G}
}

@ARTICLE{schartmann2005,
       author = {{Schartmann}, M. and {Meisenheimer}, K. and {Camenzind}, M. and {Wolf}, S. and {Henning}, Th.},
        title = "{Towards a physical model of dust tori in Active Galactic Nuclei. Radiative transfer calculations for a hydrostatic torus model}",
      journal = {\aap},
         year = 2005,
        month = jul,
       volume = {437},
       number = {3},
        pages = {861-881},
          doi = {10.1051/0004-6361:20042363},
archivePrefix = {arXiv},
       eprint = {astro-ph/0504105},
 primaryClass = {astro-ph},
       adsurl = {https://ui.adsabs.harvard.edu/abs/2005A&A...437..861S}
}

@ARTICLE{barvainis1987,
       author = {{Barvainis}, Richard},
        title = "{Hot Dust and the Near-Infrared Bump in the Continuum Spectra of Quasars and Active Galactic Nuclei}",
      journal = {\apj},
         year = 1987,
        month = sep,
       volume = {320},
        pages = {537},
          doi = {10.1086/165571},
       adsurl = {https://ui.adsabs.harvard.edu/abs/1987ApJ...320..537B}
}

@ARTICLE{asmus2014,
       author = {{Asmus}, D. and {H{\"o}nig}, S.~F. and {Gandhi}, P. and {Smette}, A. and {Duschl}, W.~J.},
        title = "{The subarcsecond mid-infrared view of local active galactic nuclei - I. The N- and Q-band imaging atlas}",
      journal = {\mnras},
         year = 2014,
        month = apr,
       volume = {439},
       number = {2},
        pages = {1648-1679},
          doi = {10.1093/mnras/stu041},
archivePrefix = {arXiv},
       eprint = {1310.2770},
 primaryClass = {astro-ph.CO},
       adsurl = {https://ui.adsabs.harvard.edu/abs/2014MNRAS.439.1648A}
}

@ARTICLE{hogbom1974,
       author = {{H{\"o}gbom}, J.~A.},
        title = "{Aperture Synthesis with a Non-Regular Distribution of Interferometer Baselines}",
      journal = {\aaps},
         year = 1974,
        month = jun,
       volume = {15},
        pages = {417},
       adsurl = {https://ui.adsabs.harvard.edu/abs/1974A&AS...15..417H}
}

@ARTICLE{lopez2022,
       author = {{Lopez}, B. and {Lagarde}, S. and {Petrov}, R.~G. and {Jaffe}, W. and {Antonelli}, P. and {Allouche}, F. and {Berio}, P. and {Matter}, A. and {Meilland}, A. and {Millour}, F. and {Robbe-Dubois}, S. and {Henning}, Th. and {Weigelt}, G. and {Glindemann}, A. and {Agocs}, T. and {Bailet}, Ch. and {Beckmann}, U. and {Bettonvil}, F. and {van Boekel}, R. and {Bourget}, P. and {Bresson}, Y. and {Bristow}, P. and {Cruzal{\`e}bes}, P. and {Eldswijk}, E. and {Fante{\"\i} Caujolle}, Y. and {Gonz{\'a}lez Herrera}, J.~C. and {Graser}, U. and {Guajardo}, P. and {Heininger}, M. and {Hofmann}, K. -H. and {Kroes}, G. and {Laun}, W. and {Lehmitz}, M. and {Leinert}, C. and {Meisenheimer}, K. and {Morel}, S. and {Neumann}, U. and {Paladini}, C. and {Percheron}, I. and {Riquelme}, M. and {Schoeller}, M. and {Stee}, Ph. and {Venema}, L. and {Woillez}, J. and {Zins}, G. and {{\'A}brah{\'a}m}, P. and {Abadie}, S. and {Abuter}, R. and {Accardo}, M. and {Adler}, T. and {Alonso}, J. and {Augereau}, J. -C. and {B{\"o}hm}, A. and {Bazin}, G. and {Beltran}, J. and {Bensberg}, A. and {Boland}, W. and {Brast}, R. and {Burtscher}, L. and {Castillo}, R. and {Chelli}, A. and {Cid}, C. and {Clausse}, J. -M. and {Connot}, C. and {Conzelmann}, R.~D. and {Danchi}, W. -C. and {Delbo}, M. and {Drevon}, J. and {Dominik}, C. and {van Duin}, A. and {Ebert}, M. and {Eisenhauer}, F. and {Flament}, S. and {Frahm}, R. and {G{\'a}mez Rosas}, V. and {Gabasch}, A. and {Gallenne}, A. and {Garces}, E. and {Girard}, P. and {Glazenborg}, A. and {Gont{\'e}}, F.~Y.~J. and {Guitton}, F. and {de Haan}, M. and {Hanenburg}, H. and {Haubois}, X. and {Hocd{\'e}}, V. and {Hogerheijde}, M. and {ter Horst}, R. and {Hron}, J. and {Hummel}, C.~A. and {Hubin}, N. and {Huerta}, R. and {Idserda}, J. and {Isbell}, J.~W. and {Ives}, D. and {Jakob}, G. and {Jask{\'o}}, A. and {Jochum}, L. and {Klarmann}, L. and {Klein}, R. and {Kragt}, J. and {Kuindersma}, S. and {Kokoulina}, E. and {Labadie}, L. and {Lacour}, S. and {Leftley}, J. and {Le Poole}, R. and {Lizon}, J. -L. and {Lopez}, M. and {Lykou}, F. and {M{\'e}rand}, A. and {Marcotto}, A. and {Mauclert}, N. and {Maurer}, T. and {Mehrgan}, L.~H. and {Meisner}, J. and {Meixner}, K. and {Mellein}, M. and {Menut}, J.~L. and {Mohr}, L. and {Mosoni}, L. and {Navarro}, R. and {Nu{\ss}baum}, E. and {Pallanca}, L. and {Pantin}, E. and {Pasquini}, L. and {Phan Duc}, T. and {Pott}, J. -U. and {Pozna}, E. and {Richichi}, A. and {Ridinger}, A. and {Rigal}, F. and {Rivinius}, Th. and {Roelfsema}, R. and {Rohloff}, R. -R. and {Rousseau}, S. and {Salabert}, D. and {Schertl}, D. and {Schuhler}, N. and {Schuil}, M. and {Shabun}, K. and {Soulain}, A. and {Stephan}, C. and {Toledo}, P. and {Tristram}, K. and {Tromp}, N. and {Vakili}, F. and {Varga}, J. and {Vinther}, J. and {Waters}, L.~B.~F.~M. and {Wittkowski}, M. and {Wolf}, S. and {Wrhel}, F. and {Yoffe}, G.},
        title = "{MATISSE, the VLTI mid-infrared imaging spectro-interferometer}",
      journal = {\aap},
         year = 2022,
        month = mar,
       volume = {659},
          eid = {A192},
        pages = {A192},
          doi = {10.1051/0004-6361/202141785},
archivePrefix = {arXiv},
       eprint = {2110.15556},
 primaryClass = {astro-ph.IM},
       adsurl = {https://ui.adsabs.harvard.edu/abs/2022A&A...659A.192L}
}

@ARTICLE{wright2010,
       author = {{Wright}, Edward L. and {Eisenhardt}, Peter R.~M. and {Mainzer}, Amy K. and {Ressler}, Michael E. and {Cutri}, Roc M. and {Jarrett}, Thomas and {Kirkpatrick}, J. Davy and {Padgett}, Deborah and {McMillan}, Robert S. and {Skrutskie}, Michael and {Stanford}, S.~A. and {Cohen}, Martin and {Walker}, Russell G. and {Mather}, John C. and {Leisawitz}, David and {Gautier}, III, Thomas N. and {McLean}, Ian and {Benford}, Dominic and {Lonsdale}, Carol J. and {Blain}, Andrew and {Mendez}, Bryan and {Irace}, William R. and {Duval}, Valerie and {Liu}, Fengchuan and {Royer}, Don and {Heinrichsen}, Ingolf and {Howard}, Joan and {Shannon}, Mark and {Kendall}, Martha and {Walsh}, Amy L. and {Larsen}, Mark and {Cardon}, Joel G. and {Schick}, Scott and {Schwalm}, Mark and {Abid}, Mohamed and {Fabinsky}, Beth and {Naes}, Larry and {Tsai}, Chao-Wei},
        title = "{The Wide-field Infrared Survey Explorer (WISE): Mission Description and Initial On-orbit Performance}",
      journal = {\aj},
         year = 2010,
        month = dec,
       volume = {140},
       number = {6},
        pages = {1868-1881},
          doi = {10.1088/0004-6256/140/6/1868},
archivePrefix = {arXiv},
       eprint = {1008.0031},
 primaryClass = {astro-ph.IM},
       adsurl = {https://ui.adsabs.harvard.edu/abs/2010AJ....140.1868W}
}

@ARTICLE{leinert2003,
       author = {{Leinert}, Ch. and {Graser}, U. and {Przygodda}, F. and {Waters}, L.~B.~F.~M. and {Perrin}, G. and {Jaffe}, W. and {Lopez}, B. and {Bakker}, E.~J. and {B{\"o}hm}, A. and {Chesneau}, O. and {Cotton}, W.~D. and {Damstra}, S. and {de Jong}, J. and {Glazenborg-Kluttig}, A.~W. and {Grimm}, B. and {Hanenburg}, H. and {Laun}, W. and {Lenzen}, R. and {Ligori}, S. and {Mathar}, R.~J. and {Meisner}, J. and {Morel}, S. and {Morr}, W. and {Neumann}, U. and {Pel}, J. -W. and {Schuller}, P. and {Rohloff}, R. -R. and {Stecklum}, B. and {Storz}, C. and {von der L{\"u}he}, O. and {Wagner}, K.},
        title = "{MIDI {\textendash} the 10 {\ensuremath{\mu}}m instrument on the VLTI}",
      journal = {\apss},
         year = 2003,
        month = aug,
       volume = {286},
       number = {1},
        pages = {73-83},
          doi = {10.1023/A:1026158127732},
       adsurl = {https://ui.adsabs.harvard.edu/abs/2003Ap&SS.286...73L}
}

@ARTICLE{matplotlib,
  author={Hunter, John D.},
  journal={Computing in Science \& Engineering}, 
  title={Matplotlib: A 2D Graphics Environment}, 
  year={2007},
  volume={9},
  number={3},
  pages={90-95},
  doi={10.1109/MCSE.2007.55}}

@ARTICLE{scikit-image,
       author = {{van der Walt}, Stefan and {Sch{\"o}nberger}, Johannes L. and {Nunez-Iglesias}, Juan and {Boulogne}, Fran{\c{c}}ois and {Warner}, Joshua D. and {Yager}, Neil and {Gouillart}, Emmanuelle and {Yu}, Tony and {scikit-image Contributors}},
        title = "{scikit-image: Image processing in Python}",
      journal = {PeerJ},
         year = 2014,
        month = jan,
       volume = {2},
          eid = {e453},
        pages = {e453},
          doi = {10.7717/peerj.453},
archivePrefix = {arXiv},
       eprint = {1407.6245},
 primaryClass = {cs.MS},
       adsurl = {https://ui.adsabs.harvard.edu/abs/2014PeerJ...2..453V}
}

@ARTICLE{numpy2020,
  author  = {Harris, Charles R. and Millman, K. Jarrod and
            van der Walt, Stéfan J and Gommers, Ralf and
            Virtanen, Pauli and Cournapeau, David and
            Wieser, Eric and Taylor, Julian and Berg, Sebastian and
            Smith, Nathaniel J. and Kern, Robert and Picus, Matti and
            Hoyer, Stephan and van Kerkwijk, Marten H. and
            Brett, Matthew and Haldane, Allan and
            Fernández del Río, Jaime and Wiebe, Mark and
            Peterson, Pearu and Gérard-Marchant, Pierre and
            Sheppard, Kevin and Reddy, Tyler and Weckesser, Warren and
            Abbasi, Hameer and Gohlke, Christoph and
            Oliphant, Travis E.},
  title   = {Array programming with {NumPy}},
  journal = {Nature},
  year    = {2020},
  volume  = {585},
  pages   = {357–362},
  doi     = {10.1038/s41586-020-2649-2}
}

@ARTICLE{scipy2020,
  author  = {Virtanen, Pauli and Gommers, Ralf and Oliphant, Travis E. and
            Haberland, Matt and Reddy, Tyler and Cournapeau, David and
            Burovski, Evgeni and Peterson, Pearu and Weckesser, Warren and
            Bright, Jonathan and {van der Walt}, St{\'e}fan J. and
            Brett, Matthew and Wilson, Joshua and Millman, K. Jarrod and
            Mayorov, Nikolay and Nelson, Andrew R. J. and Jones, Eric and
            Kern, Robert and Larson, Eric and Carey, C J and
            Polat, {\.I}lhan and Feng, Yu and Moore, Eric W. and
            {VanderPlas}, Jake and Laxalde, Denis and Perktold, Josef and
            Cimrman, Robert and Henriksen, Ian and Quintero, E. A. and
            Harris, Charles R. and Archibald, Anne M. and
            Ribeiro, Ant{\^o}nio H. and Pedregosa, Fabian and
            {van Mulbregt}, Paul and {SciPy 1.0 Contributors}},
  title   = {{{SciPy} 1.0: Fundamental Algorithms for Scientific
            Computing in Python}},
  journal = {Nature Methods},
  year    = {2020},
  volume  = {17},
  pages   = {261--272},
  adsurl  = {https://rdcu.be/b08Wh},
  doi     = {10.1038/s41592-019-0686-2},
}

@ARTICLE{astropy2022,
       author = {{Astropy Collaboration} and {Price-Whelan}, Adrian M. and {Lim}, Pey Lian and {Earl}, Nicholas and {Starkman}, Nathaniel and {Bradley}, Larry and {Shupe}, David L. and {Patil}, Aarya A. and {Corrales}, Lia and {Brasseur}, C.~E. and {N{\"o}the}, Maximilian and {Donath}, Axel and {Tollerud}, Erik and {Morris}, Brett M. and {Ginsburg}, Adam and {Vaher}, Eero and {Weaver}, Benjamin A. and {Tocknell}, James and {Jamieson}, William and {van Kerkwijk}, Marten H. and {Robitaille}, Thomas P. and {Merry}, Bruce and {Bachetti}, Matteo and {G{\"u}nther}, H. Moritz and {Aldcroft}, Thomas L. and {Alvarado-Montes}, Jaime A. and {Archibald}, Anne M. and {B{\'o}di}, Attila and {Bapat}, Shreyas and {Barentsen}, Geert and {Baz{\'a}n}, Juanjo and {Biswas}, Manish and {Boquien}, M{\'e}d{\'e}ric and {Burke}, D.~J. and {Cara}, Daria and {Cara}, Mihai and {Conroy}, Kyle E. and {Conseil}, Simon and {Craig}, Matthew W. and {Cross}, Robert M. and {Cruz}, Kelle L. and {D'Eugenio}, Francesco and {Dencheva}, Nadia and {Devillepoix}, Hadrien A.~R. and {Dietrich}, J{\"o}rg P. and {Eigenbrot}, Arthur Davis and {Erben}, Thomas and {Ferreira}, Leonardo and {Foreman-Mackey}, Daniel and {Fox}, Ryan and {Freij}, Nabil and {Garg}, Suyog and {Geda}, Robel and {Glattly}, Lauren and {Gondhalekar}, Yash and {Gordon}, Karl D. and {Grant}, David and {Greenfield}, Perry and {Groener}, Austen M. and {Guest}, Steve and {Gurovich}, Sebastian and {Handberg}, Rasmus and {Hart}, Akeem and {Hatfield-Dodds}, Zac and {Homeier}, Derek and {Hosseinzadeh}, Griffin and {Jenness}, Tim and {Jones}, Craig K. and {Joseph}, Prajwel and {Kalmbach}, J. Bryce and {Karamehmetoglu}, Emir and {Ka{\l}uszy{\'n}ski}, Miko{\l}aj and {Kelley}, Michael S.~P. and {Kern}, Nicholas and {Kerzendorf}, Wolfgang E. and {Koch}, Eric W. and {Kulumani}, Shankar and {Lee}, Antony and {Ly}, Chun and {Ma}, Zhiyuan and {MacBride}, Conor and {Maljaars}, Jakob M. and {Muna}, Demitri and {Murphy}, N.~A. and {Norman}, Henrik and {O'Steen}, Richard and {Oman}, Kyle A. and {Pacifici}, Camilla and {Pascual}, Sergio and {Pascual-Granado}, J. and {Patil}, Rohit R. and {Perren}, Gabriel I. and {Pickering}, Timothy E. and {Rastogi}, Tanuj and {Roulston}, Benjamin R. and {Ryan}, Daniel F. and {Rykoff}, Eli S. and {Sabater}, Jose and {Sakurikar}, Parikshit and {Salgado}, Jes{\'u}s and {Sanghi}, Aniket and {Saunders}, Nicholas and {Savchenko}, Volodymyr and {Schwardt}, Ludwig and {Seifert-Eckert}, Michael and {Shih}, Albert Y. and {Jain}, Anany Shrey and {Shukla}, Gyanendra and {Sick}, Jonathan and {Simpson}, Chris and {Singanamalla}, Sudheesh and {Singer}, Leo P. and {Singhal}, Jaladh and {Sinha}, Manodeep and {Sip{\H{o}}cz}, Brigitta M. and {Spitler}, Lee R. and {Stansby}, David and {Streicher}, Ole and {{\v{S}}umak}, Jani and {Swinbank}, John D. and {Taranu}, Dan S. and {Tewary}, Nikita and {Tremblay}, Grant R. and {de Val-Borro}, Miguel and {Van Kooten}, Samuel J. and {Vasovi{\'c}}, Zlatan and {Verma}, Shresth and {de Miranda Cardoso}, Jos{\'e} Vin{\'\i}cius and {Williams}, Peter K.~G. and {Wilson}, Tom J. and {Winkel}, Benjamin and {Wood-Vasey}, W.~M. and {Xue}, Rui and {Yoachim}, Peter and {Zhang}, Chen and {Zonca}, Andrea and {Astropy Project Contributors}},
        title = "{The Astropy Project: Sustaining and Growing a Community-oriented Open-source Project and the Latest Major Release (v5.0) of the Core Package}",
      journal = {\apj},
         year = 2022,
        month = aug,
       volume = {935},
       number = {2},
          eid = {167},
        pages = {167},
          doi = {10.3847/1538-4357/ac7c74},
archivePrefix = {arXiv},
       eprint = {2206.14220},
 primaryClass = {astro-ph.IM},
       adsurl = {https://ui.adsabs.harvard.edu/abs/2022ApJ...935..167A}
}

@ARTICLE{casa2022,
       author = {{CASA Team} and {Bean}, Ben and {Bhatnagar}, Sanjay and {Castro}, Sandra and {Donovan Meyer}, Jennifer and {Emonts}, Bjorn and {Garcia}, Enrique and {Garwood}, Robert and {Golap}, Kumar and {Gonzalez Villalba}, Justo and {Harris}, Pamela and {Hayashi}, Yohei and {Hoskins}, Josh and {Hsieh}, Mingyu and {Jagannathan}, Preshanth and {Kawasaki}, Wataru and {Keimpema}, Aard and {Kettenis}, Mark and {Lopez}, Jorge and {Marvil}, Joshua and {Masters}, Joseph and {McNichols}, Andrew and {Mehringer}, David and {Miel}, Renaud and {Moellenbrock}, George and {Montesino}, Federico and {Nakazato}, Takeshi and {Ott}, Juergen and {Petry}, Dirk and {Pokorny}, Martin and {Raba}, Ryan and {Rau}, Urvashi and {Schiebel}, Darrell and {Schweighart}, Neal and {Sekhar}, Srikrishna and {Shimada}, Kazuhiko and {Small}, Des and {Steeb}, Jan-Willem and {Sugimoto}, Kanako and {Suoranta}, Ville and {Tsutsumi}, Takahiro and {van Bemmel}, Ilse M. and {Verkouter}, Marjolein and {Wells}, Akeem and {Xiong}, Wei and {Szomoru}, Arpad and {Griffith}, Morgan and {Glendenning}, Brian and {Kern}, Jeff},
        title = "{CASA, the Common Astronomy Software Applications for Radio Astronomy}",
      journal = {\pasp},
         year = 2022,
        month = nov,
       volume = {134},
       number = {1041},
          eid = {114501},
        pages = {114501},
          doi = {10.1088/1538-3873/ac9642},
archivePrefix = {arXiv},
       eprint = {2210.02276},
 primaryClass = {astro-ph.IM},
       adsurl = {https://ui.adsabs.harvard.edu/abs/2022PASP..134k4501C}
}

@ARTICLE{stalevski2016,
       author = {{Stalevski}, Marko and {Ricci}, Claudio and {Ueda}, Yoshihiro and {Lira}, Paulina and {Fritz}, Jacopo and {Baes}, Maarten},
        title = "{The dust covering factor in active galactic nuclei}",
      journal = {\mnras},
         year = 2016,
        month = may,
       volume = {458},
       number = {3},
        pages = {2288-2302},
          doi = {10.1093/mnras/stw444},
archivePrefix = {arXiv},
       eprint = {1602.06954},
 primaryClass = {astro-ph.GA},
       adsurl = {https://ui.adsabs.harvard.edu/abs/2016MNRAS.458.2288S}
}

@ARTICLE{kishimoto2011,
       author = {{Kishimoto}, M. and {H{\"o}nig}, S.~F. and {Antonucci}, R. and {Millour}, F. and {Tristram}, K.~R.~W. and {Weigelt}, G.},
        title = "{Mapping the radial structure of AGN tori}",
      journal = {\aap},
         year = 2011,
        month = dec,
       volume = {536},
          eid = {A78},
        pages = {A78},
          doi = {10.1051/0004-6361/201117367},
archivePrefix = {arXiv},
       eprint = {1110.4290},
 primaryClass = {astro-ph.CO},
       adsurl = {https://ui.adsabs.harvard.edu/abs/2011A&A...536A..78K}
}

@ARTICLE{gravity2024,
       author = {{Gravity Collaboration} and {Amorim}, A. and {Bourdarot}, G. and {Brandner}, W. and {Cao}, Y. and {Cl{\'e}net}, Y. and {Davies}, R. and {de Zeeuw}, P.~T. and {Dexter}, J. and {Drescher}, A. and {Eckart}, A. and {Eisenhauer}, F. and {Fabricius}, M. and {Feuchtgruber}, H. and {F{\"o}rster Schreiber}, N.~M. and {Garcia}, P.~J.~V. and {Genzel}, R. and {Gillessen}, S. and {Gratadour}, D. and {H{\"o}nig}, S. and {Kishimoto}, M. and {Lacour}, S. and {Lutz}, D. and {Millour}, F. and {Netzer}, H. and {Ott}, T. and {Perraut}, K. and {Perrin}, G. and {Peterson}, B.~M. and {Petrucci}, P.~O. and {Pfuhl}, O. and {Prieto}, A. and {Rabien}, S. and {Rouan}, D. and {Santos}, D.~J.~D. and {Shangguan}, J. and {Shimizu}, T. and {Sternberg}, A. and {Straubmeier}, C. and {Sturm}, E. and {Tacconi}, L.~J. and {Tristram}, K.~R.~W. and {Widmann}, F. and {Woillez}, J.},
        title = "{VLTI/GRAVITY interferometric measurements of the innermost dust structure sizes around active galactic nuclei}",
      journal = {\aap},
         year = 2024,
        month = oct,
       volume = {690},
          eid = {A76},
        pages = {A76},
          doi = {10.1051/0004-6361/202450746},
archivePrefix = {arXiv},
       eprint = {2407.13458},
 primaryClass = {astro-ph.GA},
       adsurl = {https://ui.adsabs.harvard.edu/abs/2024A&A...690A..76G}
}

@ARTICLE{kishimoto2008,
       author = {{Kishimoto}, Makoto and {Antonucci}, Robert and {Blaes}, Omer and {Lawrence}, Andy and {Boisson}, Catherine and {Albrecht}, Marcus and {Leipski}, Christian},
        title = "{The characteristic blue spectra of accretion disks in quasars as uncovered in the infrared}",
      journal = {\nat},
         year = 2008,
        month = jul,
       volume = {454},
       number = {7203},
        pages = {492-494},
          doi = {10.1038/nature07114},
archivePrefix = {arXiv},
       eprint = {0807.3703},
 primaryClass = {astro-ph},
       adsurl = {https://ui.adsabs.harvard.edu/abs/2008Natur.454..492K}
}

@ARTICLE{hoenig2010,
       author = {{H{\"o}nig}, S.~F. and {Kishimoto}, M.},
        title = "{The dusty heart of nearby active galaxies. II. From clumpy torus models to physical properties of dust around AGN}",
      journal = {\aap},
         year = 2010,
        month = nov,
       volume = {523},
          eid = {A27},
        pages = {A27},
          doi = {10.1051/0004-6361/200912676},
archivePrefix = {arXiv},
       eprint = {0909.4539},
 primaryClass = {astro-ph.CO},
       adsurl = {https://ui.adsabs.harvard.edu/abs/2010A&A...523A..27H}
}

@ARTICLE{hoenig2013,
       author = {{H{\"o}nig}, S.~F. and {Kishimoto}, M. and {Tristram}, K.~R.~W. and {Prieto}, M.~A. and {Gandhi}, P. and {Asmus}, D. and {Antonucci}, R. and {Burtscher}, L. and {Duschl}, W.~J. and {Weigelt}, G.},
        title = "{Dust in the Polar Region as a Major Contributor to the Infrared Emission of Active Galactic Nuclei}",
      journal = {\apj},
         year = 2013,
        month = jul,
       volume = {771},
       number = {2},
          eid = {87},
        pages = {87},
          doi = {10.1088/0004-637X/771/2/87},
archivePrefix = {arXiv},
       eprint = {1306.4312},
 primaryClass = {astro-ph.CO},
       adsurl = {https://ui.adsabs.harvard.edu/abs/2013ApJ...771...87H}
}

@ARTICLE{hoenig2012,
       author = {{H{\"o}nig}, S.~F. and {Kishimoto}, M. and {Antonucci}, R. and {Marconi}, A. and {Prieto}, M.~A. and {Tristram}, K. and {Weigelt}, G.},
        title = "{Parsec-scale Dust Emission from the Polar Region in the Type 2 Nucleus of NGC 424}",
      journal = {\apj},
         year = 2012,
        month = aug,
       volume = {755},
       number = {2},
          eid = {149},
        pages = {149},
          doi = {10.1088/0004-637X/755/2/149},
archivePrefix = {arXiv},
       eprint = {1206.4307},
 primaryClass = {astro-ph.CO},
       adsurl = {https://ui.adsabs.harvard.edu/abs/2012ApJ...755..149H}
}

@ARTICLE{martel1998,
       author = {{Martel}, Andr{\'e} R.},
        title = "{New H{\ensuremath{\alpha}} Spectropolarimetry of NGC 4151: The Broad-Line Region-Host Connection}",
      journal = {\apj},
         year = 1998,
        month = dec,
       volume = {508},
       number = {2},
        pages = {657-663},
          doi = {10.1086/306453},
       adsurl = {https://ui.adsabs.harvard.edu/abs/1998ApJ...508..657M}
}

@ARTICLE{tsuchikawa2021,
       author = {{Tsuchikawa}, T. and {Kaneda}, H. and {Oyabu}, S. and {Kokusho}, T. and {Kobayashi}, H. and {Yamagishi}, M. and {Toba}, Y.},
        title = "{A systematic study of silicate absorption features in heavily obscured AGNs observed by Spitzer/IRS}",
      journal = {\aap},
         year = 2021,
        month = jul,
       volume = {651},
          eid = {A117},
        pages = {A117},
          doi = {10.1051/0004-6361/202140483},
archivePrefix = {arXiv},
       eprint = {2105.04792},
 primaryClass = {astro-ph.GA},
       adsurl = {https://ui.adsabs.harvard.edu/abs/2021A&A...651A.117T}
}

@ARTICLE{mor2009,
       author = {{Mor}, Rivay and {Netzer}, Hagai and {Elitzur}, Moshe},
        title = "{Dusty Structure Around Type-I Active Galactic Nuclei: Clumpy Torus Narrow-line Region and Near-nucleus Hot Dust}",
      journal = {\apj},
         year = 2009,
        month = nov,
       volume = {705},
       number = {1},
        pages = {298-313},
          doi = {10.1088/0004-637X/705/1/298},
archivePrefix = {arXiv},
       eprint = {0907.1654},
 primaryClass = {astro-ph.CO},
       adsurl = {https://ui.adsabs.harvard.edu/abs/2009ApJ...705..298M}
}

@ARTICLE{hutchings1999,
       author = {{Hutchings}, J.~B. and {Crenshaw}, D.~M. and {Danks}, A.~C. and {Gull}, T.~R. and {Kraemer}, S.~B. and {Nelson}, C.~H. and {Weistrop}, D. and {Kaiser}, M.~E. and {Joseph}, C.~L.},
        title = "{High-Velocity Line Emission in the Narrow-Line Region of NGC 4151}",
      journal = {\aj},
         year = 1999,
        month = nov,
       volume = {118},
       number = {5},
        pages = {2101-2107},
          doi = {10.1086/301076},
archivePrefix = {arXiv},
       eprint = {astro-ph/9908137},
 primaryClass = {astro-ph},
       adsurl = {https://ui.adsabs.harvard.edu/abs/1999AJ....118.2101H}
}

@ARTICLE{mundell2003,
       author = {{Mundell}, C.~G. and {Wrobel}, J.~M. and {Pedlar}, A. and {Gallimore}, J.~F.},
        title = "{The Nuclear Regions of the Seyfert Galaxy NGC 4151: Parsec-Scale H I Absorption and a Remarkable Radio Jet}",
      journal = {\apj},
         year = 2003,
        month = jan,
       volume = {583},
       number = {1},
        pages = {192-204},
          doi = {10.1086/345356},
archivePrefix = {arXiv},
       eprint = {astro-ph/0209540},
 primaryClass = {astro-ph},
       adsurl = {https://ui.adsabs.harvard.edu/abs/2003ApJ...583..192M}
}

@ARTICLE{ulvestad1998,
       author = {{Ulvestad}, James S. and {Roy}, Alan L. and {Colbert}, Edward J.~M. and {Wilson}, Andrew S.},
        title = "{A Subparsec Radio Jet or Disk in NGC 4151}",
      journal = {\apj},
         year = 1998,
        month = mar,
       volume = {496},
       number = {1},
        pages = {196-202},
          doi = {10.1086/305382},
       adsurl = {https://ui.adsabs.harvard.edu/abs/1998ApJ...496..196U}
}

@ARTICLE{leftley2024,
       author = {{Leftley}, J.~H. and {Petrov}, R. and {Moszczynski}, N. and {Vermot}, P. and {H{\"o}nig}, S.~F. and {Gamez Rosas}, V. and {Isbell}, J.~W. and {Jaffe}, W. and {Cl{\'e}net}, Y. and {Augereau}, J. -C. and {Berio}, P. and {Davies}, R.~I. and {Henning}, T. and {Lagarde}, S. and {Lopez}, B. and {Matter}, A. and {Meilland}, A. and {Millour}, F. and {Nesvadba}, N. and {Shimizu}, T.~T. and {Sturm}, E. and {Weigelt}, G.},
        title = "{Chromatically modeling the parsec-scale dusty structure in the center of NGC 1068}",
      journal = {\aap},
         year = 2024,
        month = jun,
       volume = {686},
          eid = {A204},
        pages = {A204},
          doi = {10.1051/0004-6361/202348977},
archivePrefix = {arXiv},
       eprint = {2312.12125},
 primaryClass = {astro-ph.GA},
       adsurl = {https://ui.adsabs.harvard.edu/abs/2024A&A...686A.204L}
}

@phdthesis{merritt2022,
  author  = {{Merritt}, Rachael},
  title   = "{The Spectral Energy Distributions of Active Galactic Nuclei with Direct Black Hole Mass Measurements}",
  school  = {Georgia State University},
  year    = {2022}
}

@ARTICLE{kaspi2005,
       author = {{Kaspi}, Shai and {Maoz}, Dan and {Netzer}, Hagai and {Peterson}, Bradley M. and {Vestergaard}, Marianne and {Jannuzi}, Buell T.},
        title = "{The Relationship between Luminosity and Broad-Line Region Size in Active Galactic Nuclei}",
      journal = {\apj},
         year = 2005,
        month = aug,
       volume = {629},
       number = {1},
        pages = {61-71},
          doi = {10.1086/431275},
archivePrefix = {arXiv},
       eprint = {astro-ph/0504484},
 primaryClass = {astro-ph},
       adsurl = {https://ui.adsabs.harvard.edu/abs/2005ApJ...629...61K}
}

@ARTICLE{leftley2019,
       author = {{Leftley}, James H. and {H{\"o}nig}, Sebastian F. and {Asmus}, Daniel and {Tristram}, Konrad R.~W. and {Gandhi}, Poshak and {Kishimoto}, Makoto and {Venanzi}, Marta and {Williamson}, David J.},
        title = "{Parsec-scale Dusty Winds in Active Galactic Nuclei: Evidence for Radiation Pressure Driving}",
      journal = {\apj},
         year = 2019,
        month = nov,
       volume = {886},
       number = {1},
          eid = {55},
        pages = {55},
          doi = {10.3847/1538-4357/ab4a0b},
archivePrefix = {arXiv},
       eprint = {1910.00600},
 primaryClass = {astro-ph.GA},
       adsurl = {https://ui.adsabs.harvard.edu/abs/2019ApJ...886...55L}
}

@ARTICLE{blandhawthorn1997,
       author = {{Bland-Hawthorn}, J. and {Gallimore}, J.~F. and {Tacconi}, L.~J. and {Brinks}, E. and {Baum}, S.~A. and {Antonucci}, R.~R.~J. and {Cecil}, G.~N.},
        title = "{The Ringberg Standards for NGC 1068}",
      journal = {\apss},
         year = 1997,
        month = feb,
       volume = {248},
       number = {1-2},
        pages = {9-19},
          doi = {10.1023/A:1000567831370},
       adsurl = {https://ui.adsabs.harvard.edu/abs/1997Ap&SS.248....9B}
}

@ARTICLE{li2023,
       author = {{Li}, Junyao and {Shen}, Yue},
        title = "{Constraining AGN Torus Sizes with Optical and Mid-infrared Ensemble Structure Functions}",
      journal = {\apj},
         year = 2023,
        month = jun,
       volume = {950},
       number = {2},
          eid = {122},
        pages = {122},
          doi = {10.3847/1538-4357/accade},
archivePrefix = {arXiv},
       eprint = {2302.12437},
 primaryClass = {astro-ph.GA},
       adsurl = {https://ui.adsabs.harvard.edu/abs/2023ApJ...950..122L}
}

@ARTICLE{koshida2014,
       author = {{Koshida}, Shintaro and {Minezaki}, Takeo and {Yoshii}, Yuzuru and {Kobayashi}, Yukiyasu and {Sakata}, Yu and {Sugawara}, Shota and {Enya}, Keigo and {Suganuma}, Masahiro and {Tomita}, Hiroyuki and {Aoki}, Tsutomu and {Peterson}, Bruce A.},
        title = "{Reverberation Measurements of the Inner Radius of the Dust Torus in 17 Seyfert Galaxies}",
      journal = {\apj},
         year = 2014,
        month = jun,
       volume = {788},
       number = {2},
          eid = {159},
        pages = {159},
          doi = {10.1088/0004-637X/788/2/159},
archivePrefix = {arXiv},
       eprint = {1406.2078},
 primaryClass = {astro-ph.GA},
       adsurl = {https://ui.adsabs.harvard.edu/abs/2014ApJ...788..159K}
}

@ARTICLE{may2017,
       author = {{May}, D. and {Steiner}, J.~E.},
        title = "{A two-stage outflow in NGC 1068}",
      journal = {\mnras},
         year = 2017,
        month = jul,
       volume = {469},
       number = {1},
        pages = {994-1025},
          doi = {10.1093/mnras/stx886},
archivePrefix = {arXiv},
       eprint = {2007.07932},
 primaryClass = {astro-ph.GA},
       adsurl = {https://ui.adsabs.harvard.edu/abs/2017MNRAS.469..994M}
}

@ARTICLE{ruiz2003,
       author = {{Ruiz}, M. and {Young}, S. and {Packham}, C. and {Alexander}, D.~M. and {Hough}, J.~H.},
        title = "{Near-infrared imaging polarimetry and modelling of NGC 4151}",
      journal = {\mnras},
         year = 2003,
        month = apr,
       volume = {340},
       number = {3},
        pages = {733-738},
          doi = {10.1046/j.1365-8711.2003.06239.x},
       adsurl = {https://ui.adsabs.harvard.edu/abs/2003MNRAS.340..733R}
}

@ARTICLE{lopezrodriguez2025,
       author = {{Lopez-Rodriguez}, Enrique and {Sanchez-Bermudez}, Joel and {Gonz{\'a}lez-Mart{\'\i}n}, Omaira and {Nikutta}, Robert and {Lau}, Ryan M. and {Thatte}, Deepashri and {Garc{\'\i}a-Bernete}, Ismael and {Girard}, Julien H. and {Hankins}, Matthew J.},
        title = "{JWST interferometric imaging reveals the dusty torus obscuring the supermassive black hole of Circinus galaxy}",
      journal = {Nature Communications},
         year = 2026,
        month = jan,
       volume = {17},
       number = {1},
          eid = {42},
        pages = {42},
          doi = {10.1038/s41467-025-66010-5},
archivePrefix = {arXiv},
       eprint = {2506.08077},
 primaryClass = {astro-ph.GA},
       adsurl = {https://ui.adsabs.harvard.edu/abs/2026NatCo..17...42L}
}

@ARTICLE{cornwell1983,
       author = {{Cornwell}, T.~J.},
        title = "{A method of stabilizing the clean algorithm}",
      journal = {\aap},
         year = 1983,
        month = may,
       volume = {121},
       number = {2},
        pages = {281-285},
       adsurl = {https://ui.adsabs.harvard.edu/abs/1983A&A...121..281C}
}

@INPROCEEDINGS{leisenring2012,
       author = {{Leisenring}, J.~M. and {Skrutskie}, M.~F. and {Hinz}, P.~M. and {Skemer}, A. and {Bailey}, V. and {Eisner}, J. and {Garnavich}, P. and {Hoffmann}, W.~F. and {Jones}, T. and {Kenworthy}, M. and {Kuzmenko}, P. and {Meyer}, M. and {Nelson}, M. and {Rodigas}, T.~J. and {Wilson}, J.~C. and {Vaitheeswaran}, V.},
        title = "{On-sky operations and performance of LMIRcam at the Large Binocular Telescope}",
    booktitle = {Ground-based and Airborne Instrumentation for Astronomy IV},
         year = 2012,
       editor = {{McLean}, Ian S. and {Ramsay}, Suzanne K. and {Takami}, Hideki},
       series = {Society of Photo-Optical Instrumentation Engineers (SPIE) Conference Series},
       volume = {8446},
        month = sep,
          eid = {84464F},
        pages = {84464F},
          doi = {10.1117/12.924814},
       adsurl = {https://ui.adsabs.harvard.edu/abs/2012SPIE.8446E..4FL}
}

@ARTICLE{varga2025,
       author = {{Varga}, J. and {Matter}, A. and {Millour}, F. and {Weigelt}, G. and {van Boekel}, R. and {Lopez}, B. and {Lykou}, F. and {K{\'o}sp{\'a}l}, {\'A}. and {Chen}, L. and {Boley}, P.~A. and {Wolf}, S. and {Hogerheijde}, M. and {Mo{\'o}r}, A. and {{\'A}brah{\'a}m}, P. and {Augereau}, J. -C. and {Cruz-Saenz de Miera}, F. and {Danchi}, W. -C. and {Henning}, Th. and {Juh{\'a}sz}, T. and {Priolet}, P. and {Scheuck}, M. and {Scigliuto}, J. and {van Haastere}, L. and {Zwicky}, L.},
        title = "{T CrA has a companion: First direct detection of T CrA B with VLTI/MATISSE}",
      journal = {\aap},
         year = 2025,
        month = mar,
       volume = {695},
          eid = {L21},
        pages = {L21},
          doi = {10.1051/0004-6361/202453443},
archivePrefix = {arXiv},
       eprint = {2503.08523},
 primaryClass = {astro-ph.SR},
       adsurl = {https://ui.adsabs.harvard.edu/abs/2025A&A...695L..21V}
}

@ARTICLE{may2020,
       author = {{May}, D. and {Steiner}, J.~E. and {Menezes}, R.~B. and {Williams}, D.~R.~A. and {Wang}, J.},
        title = "{The nuclear architecture of NGC 4151: on the path toward a universal outflow mechanism in light of NGC 1068}",
      journal = {\mnras},
         year = 2020,
        month = aug,
       volume = {496},
       number = {2},
        pages = {1488-1516},
          doi = {10.1093/mnras/staa1545},
archivePrefix = {arXiv},
       eprint = {2007.07374},
 primaryClass = {astro-ph.GA},
       adsurl = {https://ui.adsabs.harvard.edu/abs/2020MNRAS.496.1488M}
}

@ARTICLE{williams2020,
       author = {{Williams}, D.~R.~A. and {Baldi}, R.~D. and {McHardy}, I.~M. and {Beswick}, R.~J. and {Panessa}, F. and {May}, D. and {Mold{\'o}n}, J. and {Argo}, M.~K. and {Bruni}, G. and {Dullo}, B.~T. and {Knapen}, J.~H. and {Brinks}, E. and {Fenech}, D.~M. and {Mundell}, C.~G. and {Muxlow}, T.~W.~B. and {Pahari}, M. and {Westcott}, J.},
        title = "{The curious activity in the nucleus of NGC 4151: jet interaction causing variability?}",
      journal = {\mnras},
         year = 2020,
        month = jul,
       volume = {495},
       number = {3},
        pages = {3079-3086},
          doi = {10.1093/mnras/staa1152},
archivePrefix = {arXiv},
       eprint = {2004.10552},
 primaryClass = {astro-ph.GA},
       adsurl = {https://ui.adsabs.harvard.edu/abs/2020MNRAS.495.3079W}
}

@ARTICLE{mathis1977,
       author = {{Mathis}, J.~S. and {Rumpl}, W. and {Nordsieck}, K.~H.},
        title = "{The size distribution of interstellar grains.}",
      journal = {\apj},
         year = 1977,
        month = oct,
       volume = {217},
        pages = {425-433},
          doi = {10.1086/155591},
       adsurl = {https://ui.adsabs.harvard.edu/abs/1977ApJ...217..425M}
}

@ARTICLE{richardson1972,
       author = {{Richardson}, William Hadley},
        title = "{Bayesian-Based Iterative Method of Image Restoration}",
      journal = {Journal of the Optical Society of America (1917-1983)},
         year = 1972,
        month = jan,
       volume = {62},
       number = {1},
        pages = {55},
       adsurl = {https://ui.adsabs.harvard.edu/abs/1972JOSA...62...55R}
}

@ARTICLE{lucy1974,
       author = {{Lucy}, L.~B.},
        title = "{An iterative technique for the rectification of observed distributions}",
      journal = {\aj},
         year = 1974,
        month = jun,
       volume = {79},
        pages = {745},
          doi = {10.1086/111605},
       adsurl = {https://ui.adsabs.harvard.edu/abs/1974AJ.....79..745L}
}

@ARTICLE{cruzalebes2019,
       author = {{Cruzal{\`e}bes}, P. and {Petrov}, R.~G. and {Robbe-Dubois}, S. and {Varga}, J. and {Burtscher}, L. and {Allouche}, F. and {Berio}, P. and {Hofmann}, K. -H. and {Hron}, J. and {Jaffe}, W. and {Lagarde}, S. and {Lopez}, B. and {Matter}, A. and {Meilland}, A. and {Meisenheimer}, K. and {Millour}, F. and {Schertl}, D.},
        title = "{A catalogue of stellar diameters and fluxes for mid-infrared interferometry}",
      journal = {\mnras},
         year = 2019,
        month = dec,
       volume = {490},
       number = {3},
        pages = {3158-3176},
          doi = {10.1093/mnras/stz2803},
archivePrefix = {arXiv},
       eprint = {1910.00542},
 primaryClass = {astro-ph.SR},
       adsurl = {https://ui.adsabs.harvard.edu/abs/2019MNRAS.490.3158C}
}

@INPROCEEDINGS{hoffmann2014,
       author = {{Hoffmann}, William F. and {Hinz}, Philip M. and {Defr{\`e}re}, Denis and {Leisenring}, Jarron M. and {Skemer}, Andrew J. and {Arbo}, Paul A. and {Montoya}, Manny and {Mennesson}, Bertrand},
        title = "{Operation and performance of the mid-infrared camera, NOMIC, on the Large Binocular Telescope}",
    booktitle = {Ground-based and Airborne Instrumentation for Astronomy V},
         year = 2014,
       editor = {{Ramsay}, Suzanne K. and {McLean}, Ian S. and {Takami}, Hideki},
       series = {Society of Photo-Optical Instrumentation Engineers (SPIE) Conference Series},
       volume = {9147},
        month = jul,
          eid = {91471O},
        pages = {91471O},
          doi = {10.1117/12.2057252},
       adsurl = {https://ui.adsabs.harvard.edu/abs/2014SPIE.9147E..1OH}
}

@INPROCEEDINGS{hinz2016,
       author = {{Hinz}, P.~M. and {Defr{\`e}re}, D. and {Skemer}, A. and {Bailey}, V. and {Stone}, J. and {Spalding}, E. and {Vaz}, A. and {Pinna}, E. and {Puglisi}, A. and {Esposito}, S. and {Montoya}, M. and {Downey}, E. and {Leisenring}, J. and {Durney}, O. and {Hoffmann}, W. and {Hill}, J. and {Millan-Gabet}, R. and {Mennesson}, B. and {Danchi}, W. and {Morzinski}, K. and {Grenz}, P. and {Skrutskie}, M. and {Ertel}, S.},
        title = "{Overview of LBTI: a multipurpose facility for high spatial resolution observations}",
    booktitle = {Optical and Infrared Interferometry and Imaging V},
         year = 2016,
       editor = {{Malbet}, Fabien and {Creech-Eakman}, Michelle J. and {Tuthill}, Peter G.},
       series = {Society of Photo-Optical Instrumentation Engineers (SPIE) Conference Series},
       volume = {9907},
        month = aug,
          eid = {990704},
        pages = {990704},
          doi = {10.1117/12.2233795},
       adsurl = {https://ui.adsabs.harvard.edu/abs/2016SPIE.9907E..04H}
}

@INPROCEEDINGS{isbell2024,
       author = {{Isbell}, Jacob W. and {Ertel}, Steve and {Wagner}, Kevin and {Rousseau}, H{\'e}l{\`e}ne and {Power}, Jennifer and {Carlson}, Jared and {Becker}, Alex and {Rupert}, Justin and {Hinz}, Philip and {Hoffman}, William F. and {Leisenring}, Jarron and {Stone}, Jordan and {Spalding}, Eckhart},
        title = "{The LBTI: pioneering the ELT era}",
    booktitle = {Optical and Infrared Interferometry and Imaging IX},
         year = 2024,
       editor = {{Kammerer}, Jens and {Sallum}, Stephanie and {Sanchez-Bermudez}, Joel},
       series = {Society of Photo-Optical Instrumentation Engineers (SPIE) Conference Series},
       volume = {13095},
        month = aug,
          eid = {1309506},
        pages = {1309506},
          doi = {10.1117/12.3027270},
       adsurl = {https://ui.adsabs.harvard.edu/abs/2024SPIE13095E..06I}
}

@INPROCEEDINGS{ertel2020,
       author = {{Ertel}, Steve and {Hinz}, Philip M. and {Stone}, Jordan M. and {Vaz}, Amali and {Montoya}, Oscar M. and {West}, Grant S. and {Durney}, Olivier and {Grenz}, Paul and {Spalding}, Eckhart A. and {Leisenring}, Jarron and {Wagner}, Kevin and {Anugu}, Narsireddy and {Power}, Jennifer and {Maier}, Erin R. and {Defr{\`e}re}, Denis and {Hoffmann}, William and {Perera}, Saavidra and {Brown}, Samantha and {Skemer}, Andrew J. and {Mennesson}, Bertrand and {Kennedy}, Grant and {Downey}, Elwood and {Hill}, John and {Pinna}, Enrico and {Puglisi}, Alfio and {Rossi}, Fabio},
        title = "{Overview and prospects of the LBTI beyond the completed HOSTS survey}",
    booktitle = {Optical and Infrared Interferometry and Imaging VII},
         year = 2020,
       editor = {{Tuthill}, Peter G. and {M{\'e}rand}, Antoine and {Sallum}, Stephanie},
       series = {Society of Photo-Optical Instrumentation Engineers (SPIE) Conference Series},
       volume = {11446},
        month = dec,
          eid = {1144607},
        pages = {1144607},
          doi = {10.1117/12.2561849},
       adsurl = {https://ui.adsabs.harvard.edu/abs/2020SPIE11446E..07E}
}

@ARTICLE{yuan2020,
       author = {{Yuan}, W. and {Fausnaugh}, M.~M. and {Hoffmann}, S.~L. and {Macri}, L.~M. and {Peterson}, B.~M. and {Riess}, A.~G. and {Bentz}, M.~C. and {Brown}, J.~S. and {Dalla Bont{\`a}}, E. and {Davies}, R.~I. and {De Rosa}, G. and {Ferrarese}, L. and {Grier}, C.~J. and {Hicks}, E.~K.~S. and {Onken}, C.~A. and {Pogge}, R.~W. and {Storchi-Bergmann}, T. and {Vestergaard}, M.},
        title = "{The Cepheid Distance to the Seyfert 1 Galaxy NGC 4151}",
      journal = {\apj},
         year = 2020,
        month = oct,
       volume = {902},
       number = {1},
          eid = {26},
        pages = {26},
          doi = {10.3847/1538-4357/abb377},
archivePrefix = {arXiv},
       eprint = {2007.07888},
 primaryClass = {astro-ph.GA},
       adsurl = {https://ui.adsabs.harvard.edu/abs/2020ApJ...902...26Y}
}

@ARTICLE{isbell2021,
       author = {{Isbell}, Jacob W. and {Burtscher}, Leonard and {Asmus}, Daniel and {Pott}, J{\"o}rg-Uwe and {Couzy}, Paul and {Stalevski}, Marko and {G{\'a}mez Rosas}, Violeta and {Meisenheimer}, Klaus},
        title = "{Subarcsecond Mid-infrared View of Local Active Galactic Nuclei. IV. The L- and M-band Imaging Atlas}",
      journal = {\apj},
         year = 2021,
        month = apr,
       volume = {910},
       number = {2},
          eid = {104},
        pages = {104},
          doi = {10.3847/1538-4357/abdfd3},
archivePrefix = {arXiv},
       eprint = {2101.07006},
 primaryClass = {astro-ph.GA},
       adsurl = {https://ui.adsabs.harvard.edu/abs/2021ApJ...910..104I}
}

@ARTICLE{honig2017,
       author = {{H{\"o}nig}, Sebastian F. and {Kishimoto}, Makoto},
        title = "{Dusty Winds in Active Galactic Nuclei: Reconciling Observations with Models}",
      journal = {\apjl},
         year = 2017,
        month = apr,
       volume = {838},
       number = {2},
          eid = {L20},
        pages = {L20},
          doi = {10.3847/2041-8213/aa6838},
archivePrefix = {arXiv},
       eprint = {1703.07781},
 primaryClass = {astro-ph.GA},
       adsurl = {https://ui.adsabs.harvard.edu/abs/2017ApJ...838L..20H}
}

@ARTICLE{burtscher2013,
       author = {{Burtscher}, L. and {Meisenheimer}, K. and {Tristram}, K.~R.~W. and {Jaffe}, W. and {H{\"o}nig}, S.~F. and {Davies}, R.~I. and {Kishimoto}, M. and {Pott}, J. -U. and {R{\"o}ttgering}, H. and {Schartmann}, M. and {Weigelt}, G. and {Wolf}, S.},
        title = "{A diversity of dusty AGN tori. Data release for the VLTI/MIDI AGN Large Program and first results for 23 galaxies}",
      journal = {\aap},
         year = 2013,
        month = oct,
       volume = {558},
          eid = {A149},
        pages = {A149},
          doi = {10.1051/0004-6361/201321890},
archivePrefix = {arXiv},
       eprint = {1307.2068},
 primaryClass = {astro-ph.CO},
       adsurl = {https://ui.adsabs.harvard.edu/abs/2013A&A...558A.149B}
}

@ARTICLE{antonucci1993,
       author = {{Antonucci}, Robert},
        title = "{Unified models for active galactic nuclei and quasars.}",
      journal = {\araa},
         year = 1993,
        month = jan,
       volume = {31},
        pages = {473-521},
          doi = {10.1146/annurev.aa.31.090193.002353},
       adsurl = {https://ui.adsabs.harvard.edu/abs/1993ARA&A..31..473A}
}
\bibliographystyle{aasjournal}



\appendix
\section{Richardson-Lucy Deconvolution Results}
\begin{figure*}
    \centering
    \includegraphics[width=0.99\linewidth]{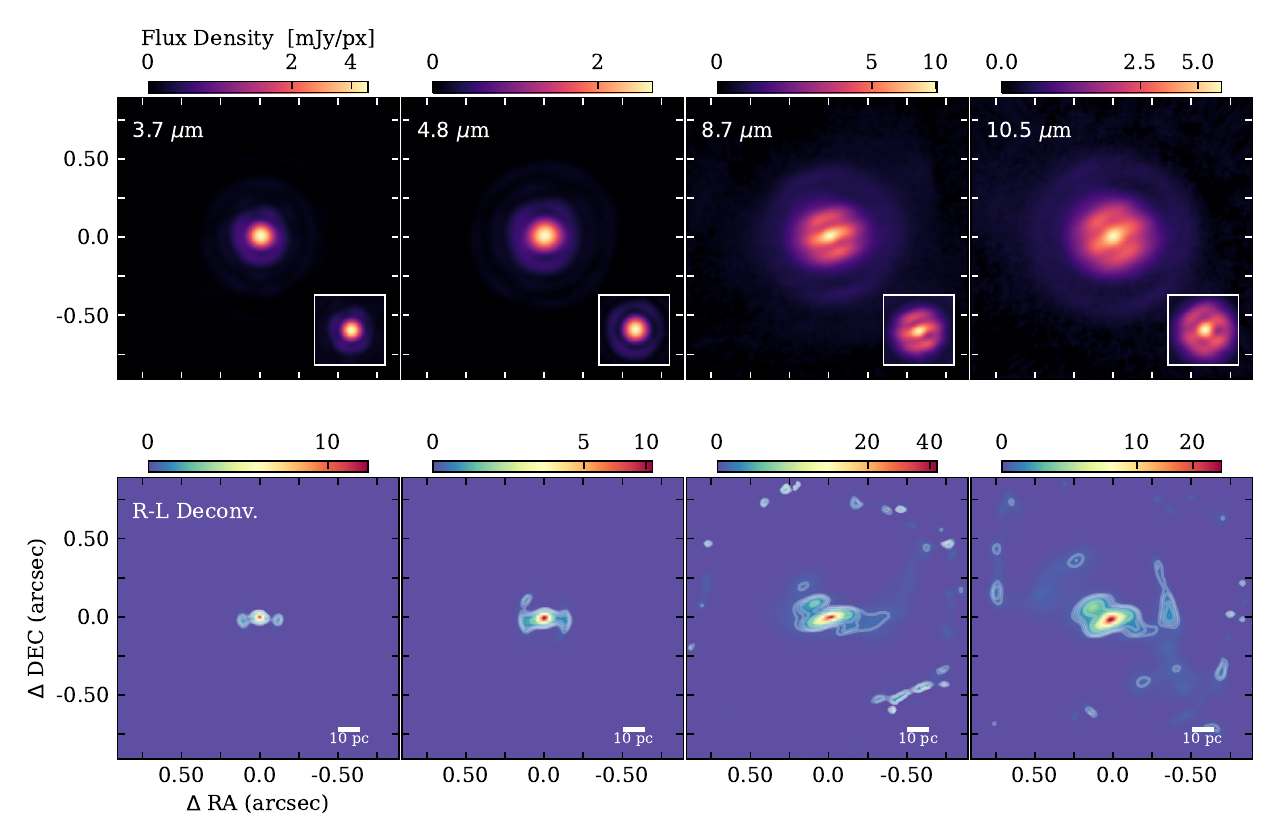}
    \caption{Deconvolved images of the nucleus of NGC 4151. \textit{Top row}) Stacked, corotated images at each wavelength (as in Figs. \ref{fig:psfs_fizeau} and \ref{fig:psfs_ao}) with PSF calibrator inset. 
    \textit{Bottom row}) R-L deconvolution results at each wavelength. 
    Contours start at $95\%$ of the peak flux and decrease by factors of 2 down to a factor of 512. 
    The LMIRCam images have been rescaled to match the pixel scale of the NOMIC images (18 mas/px).}
    \label{fig:deconv-rl}
\end{figure*}

The body of the paper focuses on the CLEAN deconvolution of the LBTI MIR images. In addition, we used the scikit-image implementation of R-L deconvolution \citep{richardson1972, lucy1974}. This required the selection of a number of iterations and a cutoff for faint features to avoid issues in division. We have balanced these values to maximize the extended features without causing obvious artifacts (such as a square feature around the edge of the image). The values are \code{niter=128} and \code{eps=1e-2}. We show the R-L images in Fig. \ref{fig:deconv-rl} to illustrate that the results are in good agreement with the CLEAN results.

\section{SED Fitting Details}
\label{app:sed_derivation}

The fitted model takes the form given in Eq. \ref{eq:bb}, which uses a positive A$_{\rm V, eff}$ to represent absorption and a negative A$_{\rm V, eff}$ to represent emission from a foreground screen. More formally, the equation for a setup involving an obscured blackbody (with temperature $T_{\rm obj}$) and a single emitting foreground screen (with temperature $T_{\rm screen}$) is
\begin{multline}
    F(T_{\rm obj}, T_{\rm screen}, A_{\rm V}) =\\B_{\lambda}(T_{\rm obj}) e^{-\tau} + (1-e^{-\tau})B_\lambda(T_{\rm screen}),
    \label{eq:full}
\end{multline}
where the first term represents the emitting object and the second term represents the foreground screen and $\tau = A_{\rm V}/1.09 \times \kappa_\lambda / \kappa_{0.5}$. Note that the circumnuclear dust is assumed to be an optically thick absorber of the incident radiation from the AGN, so we can use Kirchhoff's Law to state that the object's emissivity is approximately unity. 

We can relate Eq. \ref{eq:full} to Eq. \ref{eq:bb} by introducing an \textit{effective optical depth}, $\tau_{\rm eff}$, which can be either positive or negative,
\begin{equation}
    B_\lambda(T_{\rm obj}) e^{-\tau_{\rm eff}} = B_{\lambda}(T_{\rm obj}) e^{-\tau} + (1-e^{-\tau}) B_\lambda(T_{\rm screen}).
\end{equation}
Additionally, we assume in  Eq. \ref{eq:full} an optically thin foreground screen, so $\tau$ is taken to be small. We substitute the first-order Taylor expansion of $e^{-\tau} \approx 1-\tau$, yielding
\begin{equation}
    B_\lambda(T_{\rm obj})(1-\tau_{\rm eff}) = B_{\lambda}(T_{\rm obj})(1-\tau)  + \tau B_\lambda(T_{\rm screen}),
\end{equation}
which can be solved for $\tau_{\rm eff}$ to give 
\begin{equation}
    \tau_{\rm eff} = \tau\Big(1- \frac{B_\lambda(T_{\rm screen})}{B_{\lambda}(T_{\rm obj})}\Big) \approx \tau\Big(1- \frac{T_{\rm screen}}{T_{\rm obj}}\Big),
\end{equation}
where the latter approximation holds in the Rayleigh-Jeans regime. 
Since we use $\kappa_{\lambda}$ from \citet{schartmann2005} and relate $\tau = \frac{A_{\rm V}}{1.09}\times\frac{\kappa_\lambda}{\kappa_{0.5}}$, we finally recast this to 
\begin{equation}
    A_{\rm V,eff} = A_{\rm V} \Big(1- \frac{B_\lambda(T_{\rm screen})}{B_{\lambda}(T_{\rm obj})}\Big)
\end{equation}
to use in Eq. \ref{eq:bb}. This means that we do not fit a true extinction value, but rather an \textit{effective extinction} which is potentially wavelength dependent and can be negative if the foreground screen is bright, causing emission features (as per Eq. \ref{eq:full}). It can be thought of as the net extinction value for that line of sight. The above assumptions allow us to fit Eq. \ref{eq:bb}, which is an approximation of Eq. \ref{eq:full} with one fewer free parameter, a critical formulation for our small number of data points.

\subsection{Fit Quality}
\label{app:sed}
\begin{figure*}
    \centering
    \includegraphics[width=0.8\linewidth]{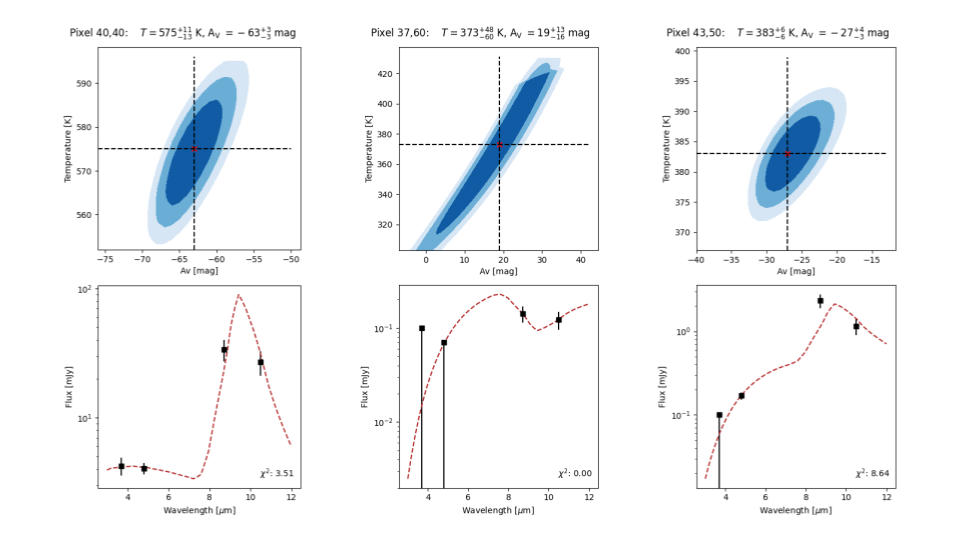}
    \caption{Example modified blackbody fits. Top row: two-dimensional $\chi^2$ contours for effective A$_{\rm V}$ and temperature. Bottom row: model spectrum (in red) compared to the observed flux measurements. Examples show three different scenarios: a fit with robust measurements, a fit with two upper limits on flux, and a fit with one upper limit on flux.} 
    \label{fig:sed-fit-examples}
\end{figure*}

\begin{figure*}
    \centering
    \includegraphics[width=0.85\linewidth]{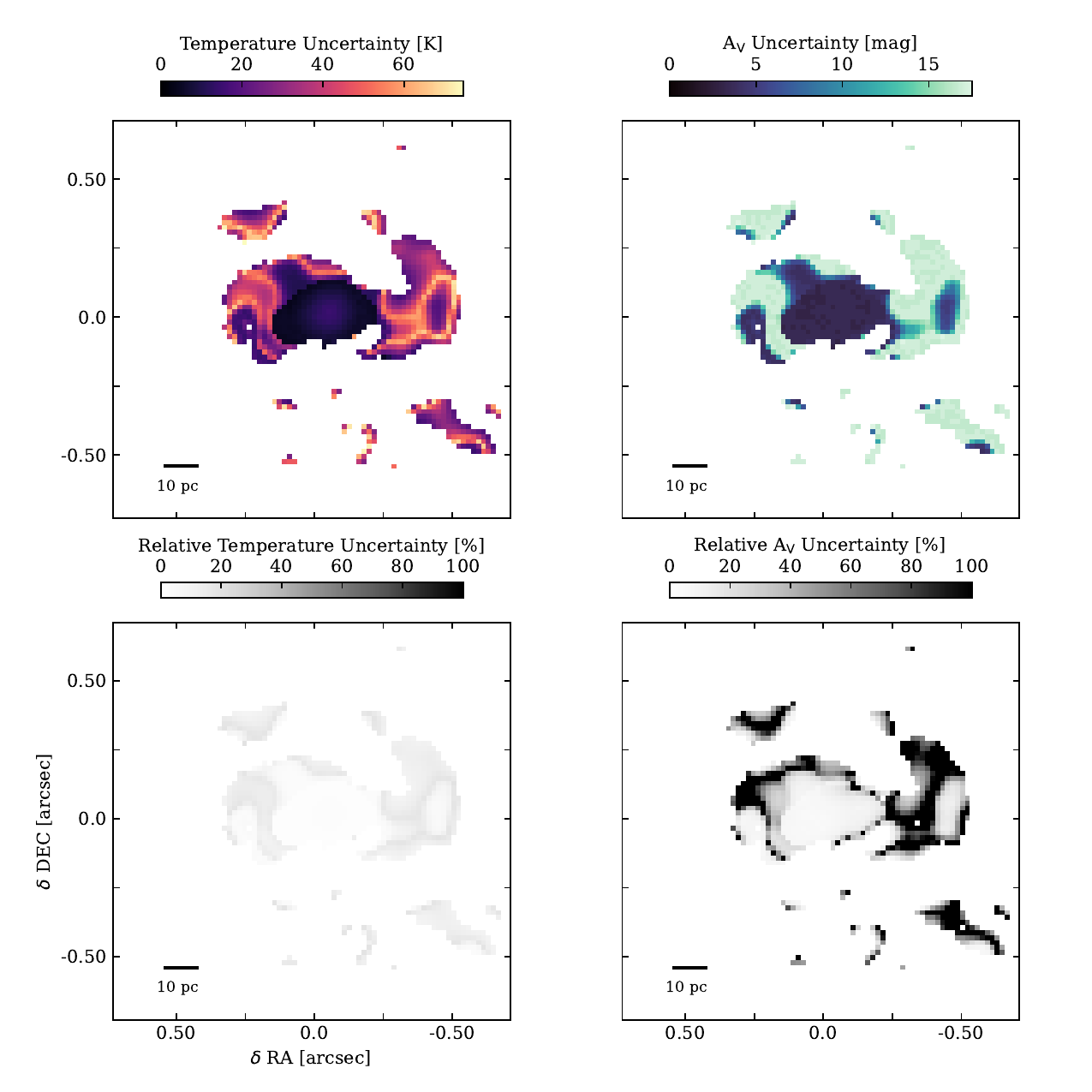}
    \caption{Modified Blackbody Fit Uncertainties. \textit{Top Row}) Absolute temperature and effective A$_{\rm V}$ uncertainty for the modified blackbody fits. \textit{Bottom Row}) Relative temperature and extinction uncertainties for the same. In the key features discussed in this work, the relative uncertainties are $\ll 10\%$. }
    \label{fig:sed_uncertainties}
\end{figure*}

While we cannot show the quality of fit for each pixel in our pixel-by-pixel SED fitting, we give several illustrative examples in Fig. \ref{fig:sed-fit-examples}. These show both the $\chi^2$ distributions of the fits as we vary temperature and extinction as well as the observations versus the model for three typical scenarios. One is when all four observations are well defined. Another is when the $LM$-fluxes are only measured as upper limits. And finally, we show the case when the $L$-band flux is given as an upper limit, but the $M$-band flux is robust. Uncertainties on the fitted parameters are derived from the $\chi^2$ distributions. Specifically, we use the $\Delta\chi^2=1$ contour to determine the $1\sigma$ confidence intervals. The resulting fitted parameter uncertainties are given in Fig. \ref{fig:sed_uncertainties}.

\begin{figure*}
    \centering
    \includegraphics[width=0.53\linewidth]{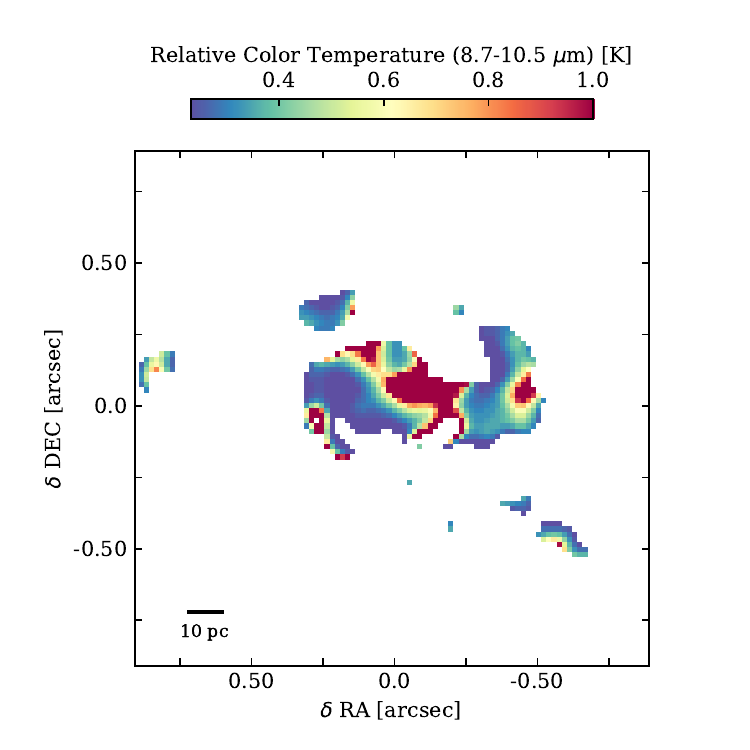}
    \caption{Model-independent color temperature estimated from the 8.7 and 10.5 $\mu$m per-pixel flux ratios. Values are normalized to the maximum temperature. The resulting relative temperature distribution is very similar to the above distributions, indicating that our results are not strongly impacted by our model assumptions. }
    \label{fig:color_temp_nband}
\end{figure*}

Finally, we also show a model-independent color temperature estimate using the flux ratios computed from the 8.7 and 10.5 $\mu$m~images in Fig. \ref{fig:color_temp_nband}. This estimate compares the flux ratio of the two wavelengths to the flux ratio of the Planck function evaluated at those same wavelengths for a number of temperatures. The resulting relative temperature distribution is very similar to the modified blackbody fits described above, and it shows that our results/interpretations are not strongly impacted by the assumptions of our modified blackbody model.

\section{VISIR Imaging Resolution}
\begin{figure*}
    \centering
    \includegraphics[width=0.75\linewidth]{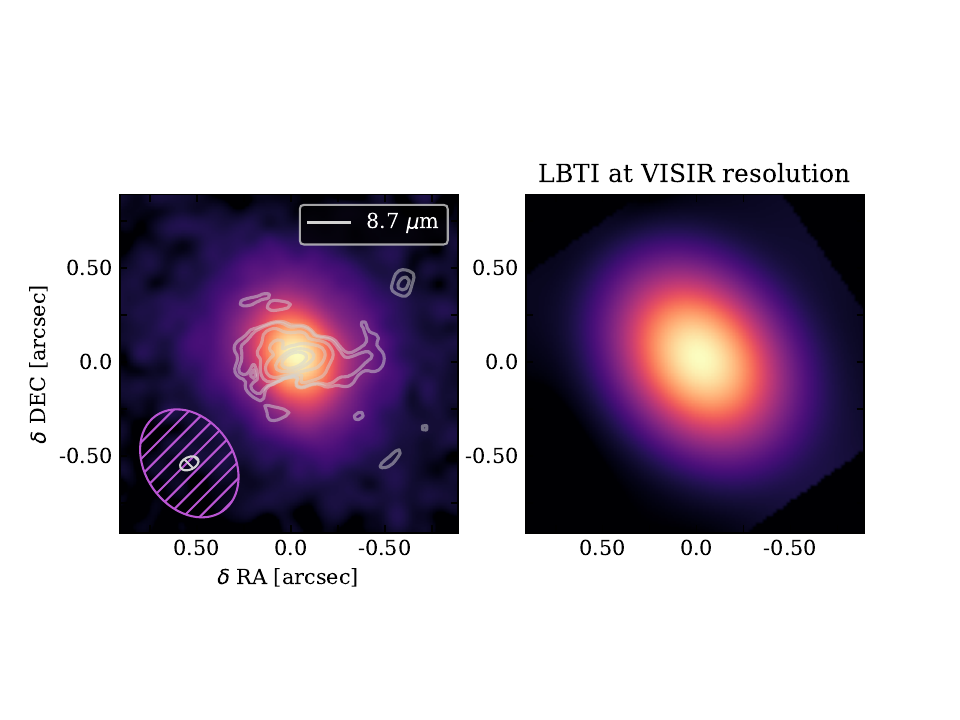}
    \caption{Comparison of results to previous VISIR imaging resolution. \textit{Left}) VISIR image with LBTI Fizeau contours overlaid. \textit{Right)} LBTI image convolved with the VISIR PSF. No E-W extension is distinguishable.}
    \label{fig:visir}
\end{figure*}

The measured $N$-band western arc was not previously detected in VISIR imaging \citep{asmus2014} at similar wavelengths despite it being at a large enough scale ($0.4$") that it would theoretically be possible. In order to understand why it was not previously detected, we convolved our $8.7$~\micron~image to the VISIR PSF FWHM reported for the previous observation ($620\times470$~mas). In the resulting image (Fig. \ref{fig:visir}), the western arc is no longer apparent. Our images are therefore not incompatible with the previous results, despite the newly detected large-scale structures. It is also important to note that the collecting area of the LBTI (using two 8.4~m mirrors) leads to higher sensitivity than for a typical VISIR snapshot.

\section{Is the Arc a Feature or an Artifact?}
\label{app:arc}
\begin{figure*}
    \centering
    \includegraphics[width=0.9\linewidth]{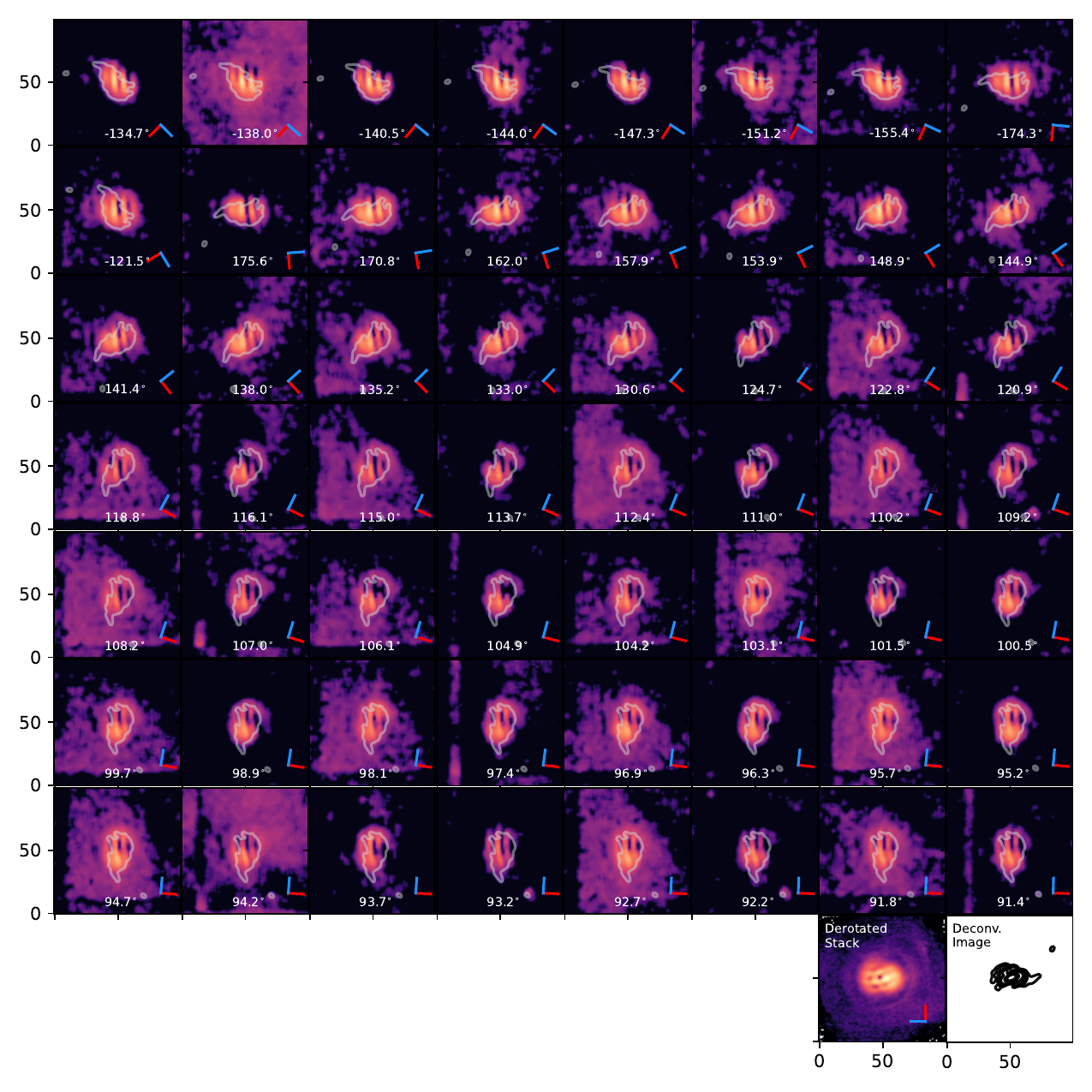}
    \caption{Mean 8.7~\micron~PSF-subtracted image of each nod position. The red line indicates North and the blue line indicates East. The CLEANed image contour is given in white for reference to various features. Images are in logarithmic scale to emphasize faint features. In the bottom panels, the stacked+derotated PSF-subtracted image and the CLEANed image contours are shown.}
    \label{fig:rot_8p7}
\end{figure*}

\begin{figure*}
    \centering
    \includegraphics[width=0.9\linewidth]{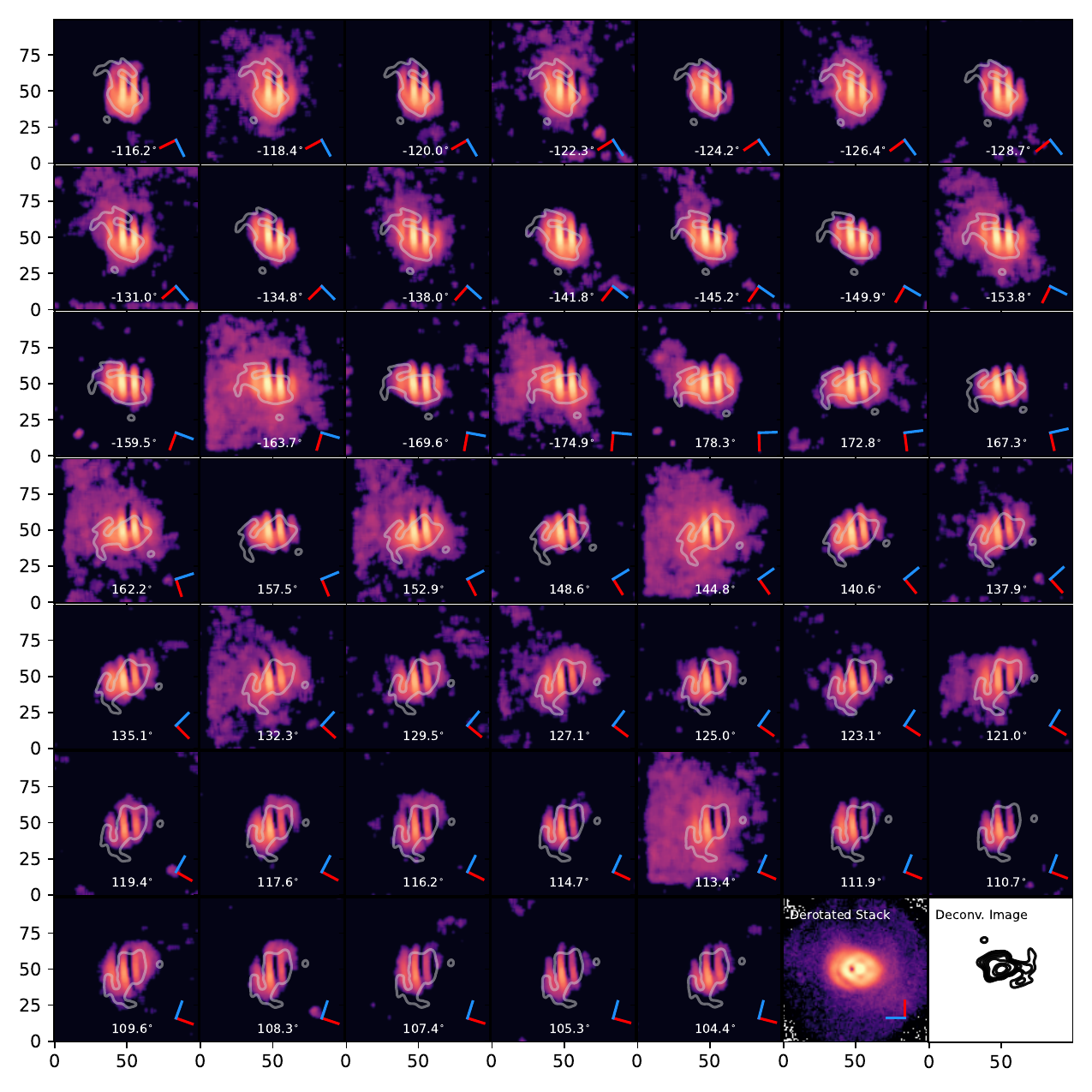}
    \caption{Mean 10.5~\micron~PSF-subtracted image of each nod position. The red line indicates North and the blue line indicates East. The CLEANed image contour is given in white for reference to various features. Images are in logarithmic scale to emphasize faint features. In the bottom right panels, the stacked+derotated PSF-subtracted image and the CLEANed image contours are shown.}
    \label{fig:rot_10p5}
\end{figure*}

The arc is thought to be robust for a few reasons:
\begin{enumerate}
\item In some of the individual PA frames (Figs. \ref{fig:rot_8p7}, \ref{fig:rot_10p5}), flux at that location is visible, and it changes with baseline rotation.
\item If it was purely an Airy ring feature, it would not be coincident in the 8.7~\micron~and 10.5~\micron~images, as the ring's diameter should change with wavelength. The arc at both wavelengths is exterior to the 8.7~\micron~Airy ring.
\item The presence of [FeII] and even H2 emission at the same location with a similar arc radius argues in favor of molecular (dusty) material being there.
\end{enumerate}
Each of these independently would not be a robust argument for its fidelity, but taken together, the Airy ring scenario is increasingly unlikely. 

\end{document}